\documentclass[12pt,preprint]{aastex} 

\received{} 
\accepted{} 
\slugcomment{Revised submission to ApJ} 
\newcommand{\ms}{\mbox{m s$^{-1}~$}} 
\newcommand{\mse}{\mbox{m s$^{-1}$}} 
\newcommand{\kms}{\mbox{km s$^{-1}~$}} 
\newcommand{\kmse}{\mbox{km s$^{-1}$}} 
\newcommand{\msun}{M$_{\odot}~$}

\newcommand{\mjup}{M$_{\rm JUP}~$}

\newcommand{\mmin}{$M_{\rm min}~$}
\newcommand{\mmine}{$M_{\rm min}$}

\begin{document} 
 
\title{Radial Velocities for 889 Late-type Stars\altaffilmark{1}}
 
\author{David L. Nidever\altaffilmark{2}, 
Geoffrey W. Marcy\altaffilmark{2,3}, 
R. Paul Butler\altaffilmark{4}, 
Debra A. Fischer\altaffilmark{3}, 
Steven S. Vogt\altaffilmark{5}} 
 
%\affil{DTM-CIW} 
%\authoraddr{} 
%\authoremail{dnidever@stars.sfsu.edu} 
\email{dnidever@stars.sfsu.edu} 
 
\altaffiltext{1}{ Based on observations obtained at the 
W.M. Keck Observatory, which is operated jointly by the 
University of California and the California Institute of Technology, 
and on observations obtained at the Lick Observatory which is 
operated by the University of California.} 
 
\altaffiltext{2}{ Department of Physics and Astronomy, San Francisco 
State University, San Francisco, CA USA 94132} 
 
\altaffiltext{3}{ Department of Astronomy, University of California, 
Berkeley, CA USA 94720} 
 
\altaffiltext{4}{ Department of Terrestrial Magnetism, Carnegie Institution 
of Washington, 5241 Broad Branch Road NW, Washington DC, USA 20015-1305} 
 
\altaffiltext{5}{ UCO/Lick Observatory, University of California at Santa Cruz, 
Santa Cruz, CA, USA 95064}

\begin{abstract} 
%We report radial velocities for 889 FGKM--type main sequence
%stars, most of which had either low-precision velocity measurements
We report radial velocities for 844 FGKM--type main sequence and subgiant
stars and 45 K giants, most of which had either low-precision velocity measurements
or none at all.  These velocities
differ from the standard stars of \cite{Udry99a} by 0.035 \kms (RMS)
for the 26 FGK standard stars in common.  The zero--point of our
velocities differs from that of Udry et al.:
$<V_{\rm present} - V_{\rm Udry} >$ = +0.053 \kmse.
Thus these new velocities agree with the best known standard stars
both in precision and zero--point, to well within 0.1 \kmse.

Nonetheless, both these velocities and the standards suffer
from three sources of systematic error, namely,
convective blueshift, gravitational redshift, and spectral type
mismatch of the reference spectrum.  These systematic errors
are here forced to be zero for G2V stars by using the Sun as reference,
with Vesta and day sky as proxies.
But for spectral types departing from solar, the systematic
errors reach 0.3 \kms in the F and K stars and 0.4 \kms in M dwarfs.

Multiple spectra were obtained for all 889 stars 
during four years, and 782 of them exhibit velocity scatter less than 
0.1 \kmse.  These stars may serve as radial velocity standards if
they  remain constant in velocity.  We found 11 new spectroscopic
binaries and report orbital parameters for them.               
\end{abstract}

%\begin{abstract} 
%We report absolute radial velocities with accuracy of 0.03 \kms for 
%889 FGKM--type main sequence stars.  The velocities agree within 0.035
%\kms (RMS) with the 26 standard stars of \cite{Udry99a}.  The
%velocity  zero--point was set using observations of the day sky and
%the asteroid  Vesta.  The resulting zero--point of our velocities is
%only slightly  different from that of Udry et al.: $<V_{\rm present} -
%V_{\rm  Udry} >$ = +0.053 \kms.  Gravitational redshifts and
%convective  blueshifts are accounted for in our velocity measurements
%to first order,  by using the known velocity of the Sun to set the
%velocity zero--point.  Nonetheless these two effects will cause
%systematic velocity errors of $\sim$0.3 \kmse as a function of
%spectral type.  Multiple spectra were obtained for all 889 stars 
%during four years, and 782 of them exhibit velocity scatter less than 
%0.1 \kmse.  These stars may serve as radial velocity standards if
%they  remain constant in velocity.  We found 11 new spectroscopic
%binaries  and report orbital parameters for them.               
%\end{abstract} 

 
%catalogs 
%techniques: radial velocities 
%techniques: spectroscopic 
%stars: fundamental parameters 
%stars: kinematics 
%stars: late-type 
%binaries: spectroscopic 
 
\section{Introduction} 
\label{intro} 
 
The radial velocity of a star is ideally the component of the
velocity vector of its center of mass that lies along the line--of--sight.  Radial 
velocities are valuable for a variety of astrophysical investigations, 
including studies of the structure of the Milky Way Galaxy, the orbits 
of long--period binary stars, and the distances to star clusters (see 
for example, \citet{bm98}).  ``Barycentric'' radial velocities (sometimes
referred to as ``absolute'' radial velocities), such as
reported here, are measured relative to the barycenter, or center of mass,
of the Solar System.  Such velocities are often (incorrectly) termed 
``heliocentric'', though the Sun moves with a speed of $\sim$13 \ms 
relative to the barycenter. 

%  ``Absolute'' radial velocities are 
%measured relative to an inertial frame such as the barycenter of the 
%Solar System.  Such velocities are often (incorrectly) termed 
%``heliocentric'', though the Sun moves with a speed of $\sim$13 \ms 
%relative to the barycenter. 
 
Radial velocities of stars in the Galaxy are often measured 
with an accuracy of only $\sim$0.5 \kmse.  With advances in the 
accuracy of proper motion measurements to $\sim$1 mas/yr (e.g., 
Perryman et al. 1996) for many stars, a corresponding increase in the 
accuracy of radial velocities is required.  Meanwhile, the best {\em 
relative} radial velocities have precisions of 3 \ms  
\citep{But00} and new instruments (e.g. HARPS) are 
designed to achieve a precision of 1 \ms \citep{Bou01, pepe00}.   
These relative velocities have proved useful in the detection of extrasolar 
planets (e.g., \citet{MCM00}) but they are not necessarily tied to a 
velocity zero--point.  Nonetheless, the precision--velocity 
instruments have overcome many observational and technical hurdles 
related to spectroscopic Doppler--shift measurements, either by using 
a gas absorption cell or a fiber--fed comparison lamp spectrum 
\citep{Val95, But96, Bar00}. 
 
The precision--velocity technology has been applied to the 
establishment of barycentric radial velocities, most notably by the 
Geneva team (Udry et al. 1999a,b).  They have measured 
velocities for 38 stable dwarf stars with precision better than 0.05 \kmse. 
 
Here we provide barycentric radial velocities with typical accuracies
of 0.3 \kms (and precise to 0.03 \kms for a given spectral type) for
889 stars.  Our intent is to provide a velocity measurement at the
current epoch for a variety of purposes.  Velocity variations with a
time scale of hundreds of years may be detected by comparison of
present and future velocities.  We also hope to establish radial
velocity standard stars, by indentifying a subset that exhibit no
significant velocity variation above 10 \mse.

\section{Barycentric Radial Velocities} 
 
The Doppler searches for planets have been successful because of the 
relative ease with which the change in radial velocity may be measured 
with high internal precision.  Such relative velocities circumvent the 
technical challenges associated with the determination of an accurate 
velocity zero--point and they avoid the physical 
interpretation of a barycentric Doppler shift which becomes
% an ``absolute'' Doppler shift which becomes 
bewildering at levels below 1 \kmse.   
 
Barycentric Doppler shifts carry an unclear interpretation for several 
reasons.  Stellar lines suffer a gravitational redshift \citep{MTW73} 
upon leaving the stellar photosphere, yielding effective redshifts of 
$V_{\rm grav} = GM/Rc$ \citep{DGL99}.  This redshift varies from 680 
\ms for F5V to 500 \ms for M5V stars, the range of spectral types 
considered here.  Uncertainties in stellar masses and radii prevent an 
accurate removal of this effect.  Furthermore, stellar lines suffer a 
transverse Doppler effect (essentially time dilation), the removal of 
which requires knowledge of the full velocity vector of the star's 
space motion.  This effect is $\simeq$50 \ms for the fastest moving 
stars \citep{Lin99}. 
 
More importantly, stellar Doppler shifts are affected by 
sub--photospheric convection (granulation), macro-turbulence, stellar 
rotation, pressure shifts, oscillations and activity cycles.  The most 
important of these effects is granulation.  The textbook explanation 
is that a larger contribution to the stellar flux emerges from hot, rising 
gas than from falling gas in the convective cells.  These motions yield
an overall blueshift of spectral lines \citep{Dra99}.  However, the 
exact convective blueshift depends on the full 3--D hydrodynamics and 
radiative transfer in each spectral line  
as a function of depth in the photosphere 
\citep{Dra99}.  The blueshift clearly depends on spectral type and is 
expected to be $\sim$--1000 \ms for F5V, --400 \ms for G2V, and --200 
\ms for K0V \citep{DGL99}.  Effects due to pressure shifts are less 
than 100 \ms for ordinary stars \citep{DGL99, APr97}.  Stellar 
rotation also imposes minor radial velocity effects \citep{Gra99}. 
 
Moreover, the measurement of spectroscopic barycentric radial
velocities usually requires a reference stellar spectrum.  Due to
constraints of telescope time and available standard stars, only a
small number of reference spectra can be used.  Typically, the Doppler
measurements require use of reference spectra having different
spectral types than the program star, which leads to spectral mismatch
errors.  Since the strengths of spectral lines vary with temperature
and metalicity, the spectra of the reference and program stars will be
significantly different.  In effect, the relative displacement in
wavelength between two non--identical spectra is not uniquely defined
and therefore is dependent on the algorithm used.  Such spurious
Doppler effects are minimized, but not eliminated, by using
high--resolution spectra with many lines resolved, as adopted in this
present survey.
 
The effects of convection, stellar gravitational redshift, transverse
Doppler shift, and the other effects mentioned above cannot be
determined for a star with an accuracy that is comparable to the
internal measurement errors of $\sim$10 \ms.  Removal of such effects
by a model of each star might introduce more model--dependent errors.
Nonetheless, with considerable modeling, the measurement of a Doppler
shift may be used to determine the ``true'' velocity component of the
center of mass of the star.  Alternatively (and more traditionally)
Doppler measurements may be left merely as an observable, namely the
spectroscopic shift in wavelength, often quoted as $z =
\Delta\lambda/\lambda_0$ \citep{Lin99}.
 
Here, we adopt the philosophy that spectroscopic barycentric radial
velocities should first be corrected for local effects, such as that
caused by the observer's motion relative to the Solar System
barycenter ($\sim$30 \kmse) and by the solar gravitational blueshift
($\sim$3 \mse).  In addition, we will correct all of our velocities of
FGK stars for gravitational redshift and convective blueshift to first
order by using the known radial velocity of the Sun to set the
zero--point for the stellar velocity measurements.
%For M stars we adopted the velocity zero--point of \citep{Mar97}.

Our quoted Doppler shifts represent  
velocities as if measured at the Solar System barycenter but with the 
Sun and its potential well removed.   
Clearly the velocity measurements 
presented here are amenable to future corrections for the spectral--type dependence
relative to G2V for gravitational redshift and convective blueshift, in order to yield 
the most accurate velocities possible for them (e.g., \citet{Gul99,SF00}).

\section{Spectroscopic Observations} 
 
The spectra were obtained with the HIRES echelle spectrometer 
\citep{Vog94} on the 10-m Keck 1 telescope and with the ``Hamilton'' 
echelle spectrometer fed by either the 3-m Shane or the 0.6-m Coude 
Auxilliary (CAT) Telescopes \citep{Vog87}.  During an observation the 
starlight is sent through a glass cell that is filled with iodine 
vapor \citep{Mar92} before entering the spectrometer, which 
superimposes iodine lines on the stellar lines.  These iodine 
absorption lines are used to calibrate the wavelength scale of the 
spectrum from 5000 \AA -- 6000 \AA. 
 
Our Doppler planet search project contains 889 stars
at the Keck and Lick Observatories \citep{But00}.  Until 
now only relative radial velocities have been computed from these 
spectra and they have a precision of $\sim$3 \ms \citep{But96, Vog00} 
which has allowed the discovery of Jovian and sub-Jovian sized 
extrasolar planets \citep{MCM00}. 
 
To achieve barycentric velocities, we adopt two standard spectra.  We use
the National Solar Observatory (NSO) FTS solar spectrum \citep{Kur84}
and an M dwarf composite spectrum (see \S4) as reference spectra.  The
889 stars each typically have $\sim$12 spectra obtained during four years
from 1997 to 2001.  The distribution of spectral types in our sample
is: 14\% F, 46\% G, 27\% K, and 13\% M stars.  Except for 45 K giants
all stars are main--sequence dwarfs or subgiants.  All stars are void of a
visible companion within 2 arcsec, though some were subsequently
revealed to be spectroscopic binaries (see \S 9).

\section{Doppler Method} 
 
The barycentric radial velocities reported here are found in a manner 
similar to that used to find the relative radial velocities for the 
planet search.  An observed spectrum is fit with a synthetic 
spectrum that is composed of the individual stellar and iodine spectra.   
In detail, the synthetic spectrum is the product of the 
deconvolved stellar ``template'' spectrum (with the spectrometer instrumental profile 
removed) with a high resolution spectrum of molecular iodine. 
The product of these two is convolved with the instrumental profile 
of the spectrometer (at the time of the observation), to produce the final synthetic spectrum, 
as described in \citet{But00}. 
 
The observed spectrum to be synthesized is broken into "chunks" of 
length 40 pixels corresponding to roughly $\sim$2 \AA.  In total 14 
free parameters are fit in the Doppler analysis; 11 devoted to the 
instrumental profile, along with the wavelength zero--point of each 
chunk, the wavelength dispersion across each chunk, and the Doppler 
shift of the stellar spectrum relative to the stellar template of
that star, $z$ ($=\Delta\lambda/\lambda$).  All of 
the parameters, except for $z$, are extracted primarily from the 
iodine portion of the observed spectrum.  After the best parameters 
are found for all the chunks of a spectrum they are saved for further 
analysis, notably the weighted average of $z$.  We apply a correction 
to all velocities for our topocentric motion relative to the 
barycenter \citep{Chris95}.  For a more in depth discussion of this 
standard Doppler analysis to obtain relative velocities, see 
\citet{Mar92,But96,Val95}. 
 
To obtain barycentric radial velocities, the approach was similar to the 
standard analysis described above and indeed we used some parameters derived from that 
analysis.  Here, we used the National Solar Observatory (NSO) solar 
spectrum \citep{Kur84} as the deconvolved template, for all F, G and K stars.
%instead of the individual stellar templates as in the planet search.
For 
M dwarfs, we constructed an in--house M star composite spectrum as the 
template (see below).  Since all parameters except for $z$ are extracted from the 
iodine portion of the spectrum previously, the use of a different 
stellar template spectrum does not affect these parameters 
significantly.  Therefore in our new fit, we simply adopt the values 
of the 13 non--$z$ parameters that were previously obtained in the 
Doppler analysis for relative velocities (i.e., for the planet search) 
and we fit here only for the barycentric radial velocity, $z$.
 
Ordinarily, when the deconvolved stellar spectrum of the individual
star is used as the template, the
%stellar template is used as the reference, the 
$\chi_{\nu}^2$ statistic is near unity (usually less than 1.7), 
because the stellar contribution to the model is a previously obtained 
deconvolved spectrum of the same star.  But in our fit for barycentric
velocities, the value of reduced $\chi_\nu^2$ is much larger (poorer) 
because of the mismatch between the template spectrum (NSO or M dwarf 
composite) and that of the program star.  The program star differs 
from the Solar or M--dwarf template in spectral type, metalicity, and 
$v\sin i$.  Thus, the spectral fits are not as good, yielding 
$\chi_\nu^2$ of 3--7. 
 
The internal error per observation, as defined by the weighted standard 
deviation of the mean of the velocities from all ($\sim$400) chunks, 
is higher for our barycentric radial velocities than for the relative 
radial velocities used in the planet search.
% We quote these large uncertainties as a lower limit to the errors in
%our absolute velocities. 
We find that while the relative velocities carry errors 
of $\sim$3 \ms (for the planet search) the average internal error per 
observation for the barycentric velocities reported here is $\sim$20 
\mse.  A more conservative estimate of our velocity precision
is given in \S6 as 0.03 \kms from 
comparison with the standard stars. 
 
\begin{figure} 
\plotone{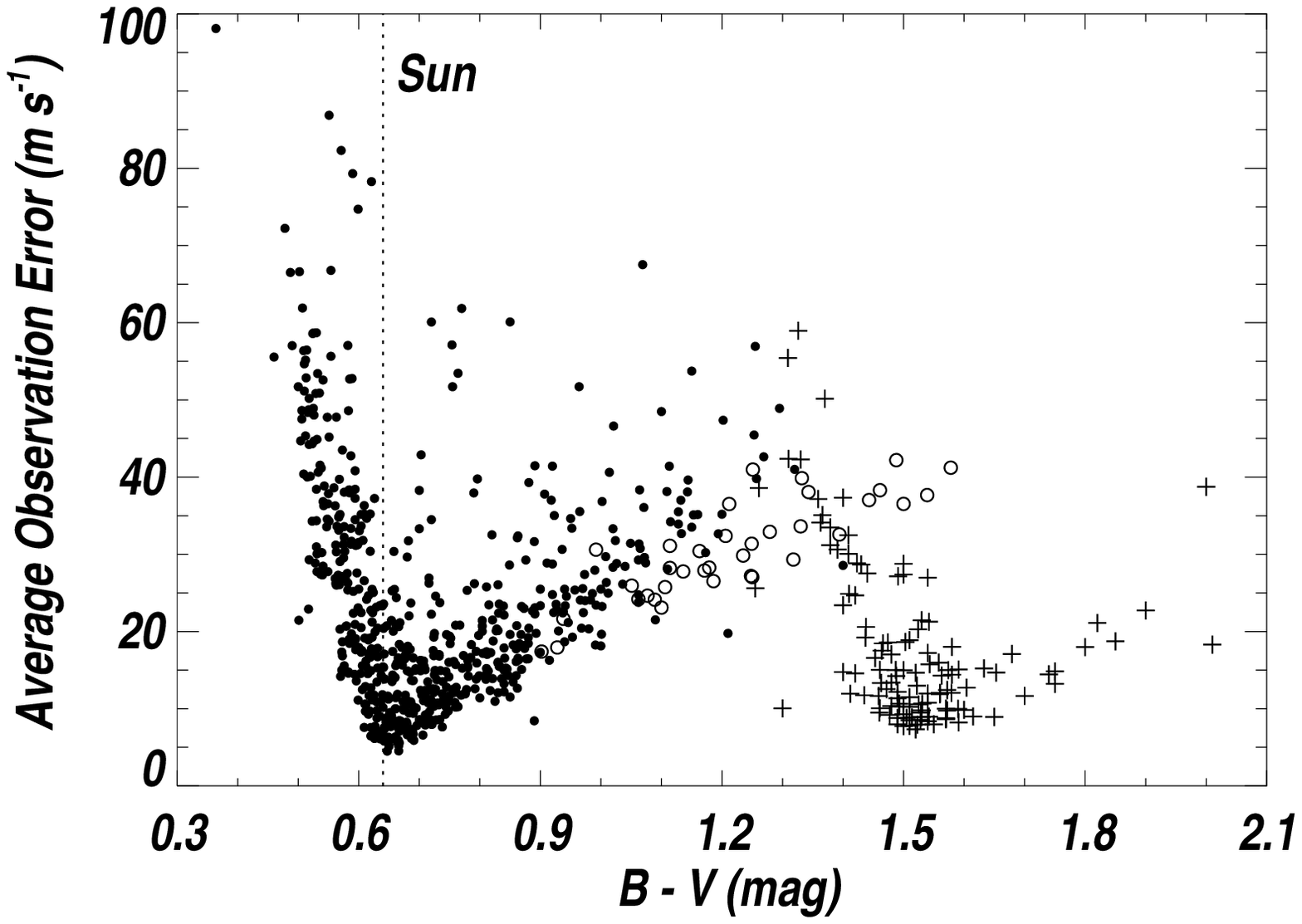} 
\caption{The internal velocity error per observation (averaged per star) vs. \bv.  The 
dots represent the stars for which the NSO solar spectrum was used as 
the reference (filled dots represent dwarfs and open dots represent giants) and the
plusses represent the stars for which the 
M star composite spectrum was used as the reference.  The \bv \ for the 
Sun (\bv = 0.64) is shown for clarity.  The spectral type dependence 
of errors is apparent.} 
\label{err_bv} 
\end{figure}

Due to the fact that only one stellar template is used for a large 
range of stellar types, we expect that our errors (internal error per observation)
will depend on \bv . 
The greater the difference between template and program star, the 
greater the expected error due to spectral mismatch.  We plot this 
internal error in our barycentric velocities versus \bv \ in Figure 
\ref{err_bv}.  Since the NSO solar spectrum, \bv = 0.64 \citep{Car96}, 
is used for F, G and K stars we expect the errors to be minimized for 
solar type stars.  This is confirmed in Figure \ref{err_bv} .  The M 
dwarf composite template spectrum was created from five different M 
dwarfs having average spectral type of M3 (see below).  The internal velocity 
errors for stars analyzed with this stellar template are minimized at 
\bv $\approx$ 1.5, as can also be seen in Figure \ref{err_bv}. 
 
Each program star has an average of 12 observations.  As many 
observations as possible are analyzed per star, up to 30, in order to 
minimize the uncertainty in the mean of the barycentric radial 
velocities.  If the radial velocity of the star were stable we should 
obtain an uncertainty in the mean of $\sim$20/$\sqrt{12} \approx$ 6 
\mse.  A majority of stars indeed exhibit velocity scatter of $\sim$10 
\ms (RMS) while others have an RMS scatter larger than 100 \mse. 
Scatter in the latter stars is almost always caused by companions or 
large chromospheric activity.  Overall, 782 of the 889 stars have an RMS velocity 
scatter less than 0.1 \kmse, and 107 stars have an RMS velocity scatter 
larger than 0.1 \kmse.  The barycentric radial velocities for these stars 
are reported in Tables 1 and 2 respectively.
 
%\subsection{M Star Composite Spectrum} 
 
As mentioned above, the NSO solar spectrum could not be used as the 
template for the M dwarf program stars because the spectra are too 
different.  Therefore, a separate reference spectrum was required for 
the M stars.  For this purpose a composite spectrum of five M stars 
was produced in the following manner.
 
Five M stars were selected having barycentric radial velocities reported 
by \citet{Mar87} with low uncertainties: GJ~251 (M4), GJ~411 (M2), 
GJ~526 (M4), GJ~752A (M3.5), and GJ~908 (M2).  One spectrum of high S/N 
ratio was used from each of these M dwarfs.  The spectra were shifted 
back, to remove the Doppler shifts caused by the barycentric motion of 
the observatory and by the barycentric radial velocity of the star itself 
relative to the barycenter, taken from \citet{Mar87}.  To check for any residual 
Doppler shifts, these corrected spectra were then cross-correlated 
with respect to one of them, GJ~251.  Assuming the resulting 
displacements were due to random errors, the mean of the residual 
velocities was taken to be the barycentric velocity zero-point.  Using 
this reference point the spectra were again corrected for their 
doppler shifts.  Finally, the spectra were put on the same wavelength 
scale and coadded to create a M star composite spectrum. 
 
\section{Velocity Zero-Point} 
 
Using observations of the day sky and the minor planet Vesta we found 
the zero-point of our velocities for FGK stars.  These references were used because 
they have essentially solar spectra, and 
the radial velocities of Vesta and the sun relative to a topocentric 
observer are easily determined.  We used the online ``JPL Ephemeris 
Generator'' to find these instantaneous velocities 
\footnote{http://ssd.jpl.nasa.gov/cgi-bin/eph}.   
Four observations of the day sky and two observations of Vesta were
analysed with the before mentioned Doppler method using the NSO solar spectrum.
These references showed that 
our raw velocities, from Keck and Lick, were consistently large by 
522$\pm$5 \ms.   
 
There are various possible sources for the 522 \ms offset in our raw 
velocities.  The absolute wavelength scales of both the NSO solar 
spectrum and our FTS iodine spectrum are not well known.  According to 
\citet{Kur84}, the solar lines are broadened by 200 \ms due to the 
change in the radial velocity during the observation, and the 
wavelengths may have errors as large as 100 \ms.  The wavelength scale 
of the iodine FTS spectrum comes from the calibration made at the 
McMath telescope at Kitt Peak.  We have no independent way to verify 
the integrity of the zero--point in wavelength of this FTS iodine 
spectrum.  The third concern stems from the instrumental profile of 
the HIRES and the Hamilton spectrometers.  It is known that the PSF is 
asymmetric to some degree \citep{Val95} and may give rise to 
systematic velocity shifts.  However, it is unlikely that the same 
asymmetry in the same direction would be found at two different 
spectrometers.  Therefore, we consider this possibility less 
plausible. 
 
All Doppler measurements which used the NSO solar spectrum as 
the reference template were corrected for this offset of 522 \mse.   
Since currently no M dwarfs exist having a definitive barycentric radial 
velocity, the velocity zero-point for the M stars in our sample 
was set using the previously published velocities for the five standards 
from \citet{Mar87} which have errors of $\sim$0.4 \kmse.

\section{Comparison of Present Velocities with Standard Stars} 
 
To get an external measure of the accuracy of our barycentric radial 
velocities we have compared our velocities to 
published velocities of supposed radial velocity standard stars.   
We carry out this comparison separately for the FGK stars and for the 
M stars. 
 
\subsection{F, G and K Stars} 
 
\begin{figure} 
\plotone{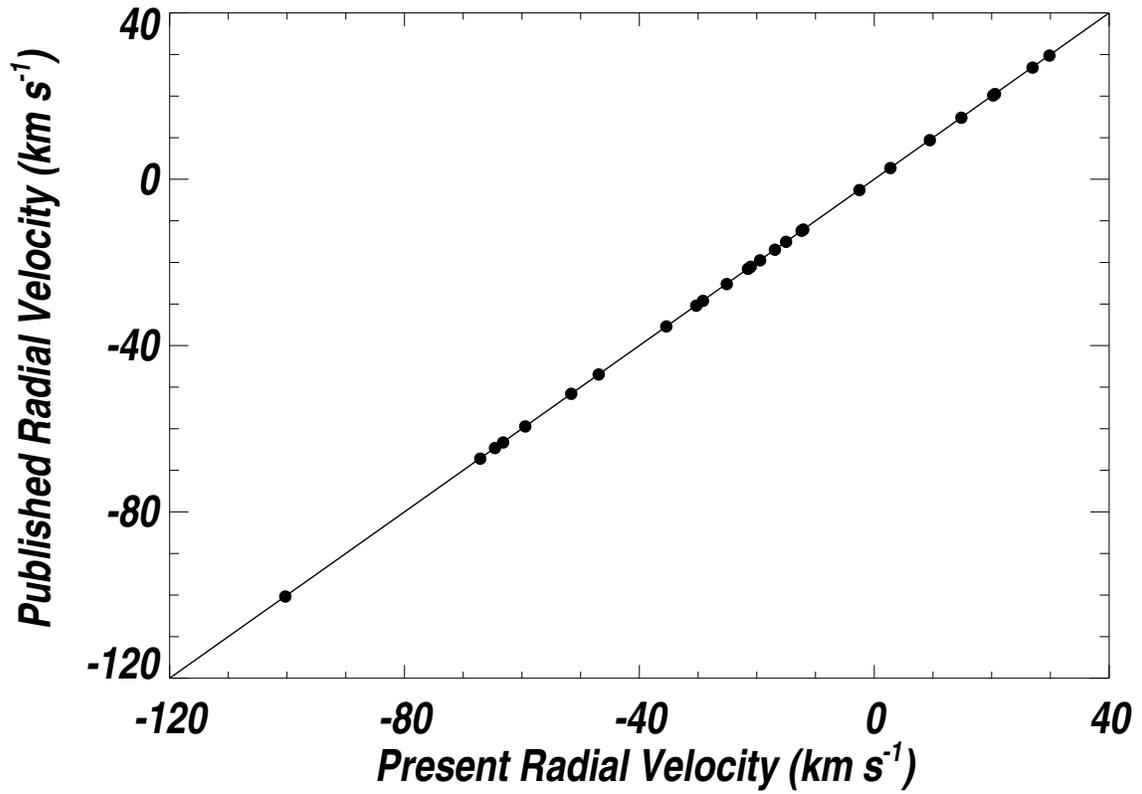} 
\caption{Velocities of standard stars vs. velocities measured here, 
for F, G, and K stars.  The standard stars are on the CORAVEL  
scale \citep{Udry99a}.  The present velocities agree well with the 
standards, with no visible nonlinear departure of velocity scale.} 
\label{cor_lin} 
\end{figure}

\begin{figure} 
\plotone{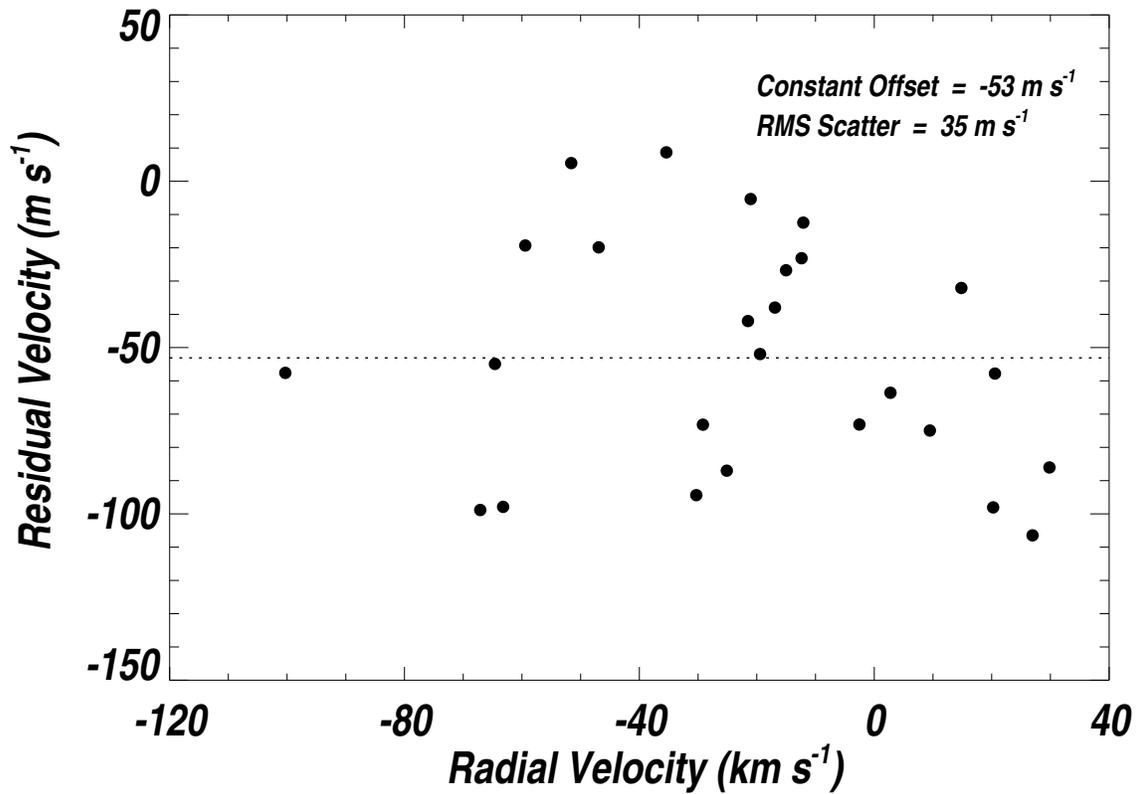} 
\caption{Standard--star velocities \citep{Udry99a}  
minus present velocities for all 26 FGK stars in common (as in 
Figure 2).  The differences reveal that the present velocities 
are higher than those of Udry et al. by 53 \mse,  
and exhibit an rms scatter of 35 \mse. Thus the 
present velocities and those of \citet{Udry99a} each have internal  
accuracy better than 35 \mse, and differ in zero--point by $\sim$53 \mse.} 
\label{cor_res} 
\end{figure}

We compared our velocities to those of the 26 standard stars 
that were measured by \citet{Udry99a}.   
The results of this comparison are shown in Figures 
\ref{cor_lin} and \ref{cor_res} and yield  
$<V_{Udry} - V_{Present}>$ = $-53$ \mse, with an rms scatter of 35 \mse.   
The corresponding uncertainty in the mean is 
35/$\sqrt{26}$ = 7 \ms . 
Thus the formal difference in zero points is 
 
\begin{equation} 
<V_{\rm Present} - V_{\rm Udry}> = +53 \pm 7 \mse 
\end{equation} 
 
\noindent 
 
Thus, there appears to be a statistically significant difference in the 
zero--point of the velocities reported here compared to those of 
\citet{Udry99a} of 53 \mse.  We do not know the origin of this 
difference, nor do we know which scale is more ``accurate''. Few 
studies will be affected by such small differences.  But future 
highly precise proper motion measurements and precise orbit 
calculations may require such accurate velocities, including 
proper treatment of gravitational redshift and convective 
blueshift. 
 
\begin{figure} 
\plotone{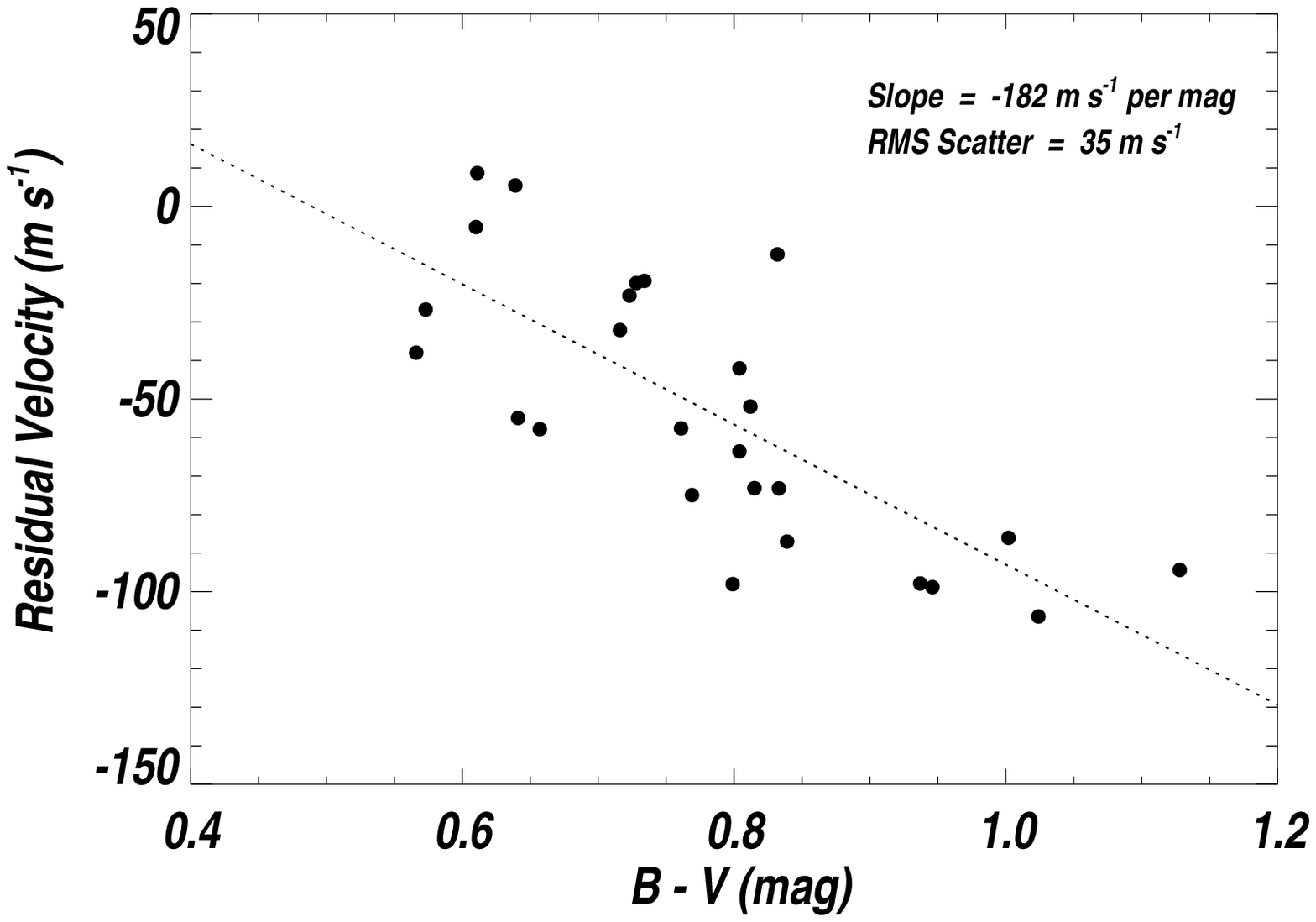} 
\caption{Standard--star velocities \citep{Udry99a} minus the present velocities
as a function of \bv.  A dependence is apparent, suggesting systematic 
errors in at least one set of velocities.  The slope is $-182 \pm 24$ \ms 
per mag.  The RMS scatter is 35 \ms before fitting and 23 \ms after
fitting a line to the data.} 
\label{cor_bv} 
\end{figure}

However, the difference between the present velocities and those of 
\citet{Udry99a} do exhibit a significant \bv \ dependence as seen in 
Figure \ref{cor_bv}.  The slope of the linear trend is $-182 \pm 24$ 
\ms per mag with an RMS scatter around the fit of 23 \mse.
This dependence is similar to the \bv \ dependence seen between 
the CfA and CORAVEL data \citep{Stef99,Udry99a}.  This color 
dependence is likely caused by some spectral type mismatch in one or 
all of the radial--velocity scales.  For solar type stars (near G2V)  
where our zero-point is well set,  
we are in good agreement with the CORAVEL 
velocities, with an offset of only 25 \ms as seen in Figure 
\ref{cor_bv}. 
 
According to \citet{Udry99a} the ELODIE measurements ensure temporal 
stability of better than 15 \ms during time scales of years  
for their standard stars .  However, their quoted 
velocities have been rounded off at the 50 \ms level.  From our 
measurements of these stars in common, the temporal variability is 
less than 10 \ms during $\sim$5 yr.   
Therefore it is not certain whether the rms 
scatter of 35 \ms between the two sets of velocities is due to our 
errors or the rounding of the ELODIE data plus their errors.  We 
conclude that the barycentric velocities reside on the same scale with an 
offset of $\sim$50 \mse, and a scatter of $\sim$35 \mse.
 
\begin{figure} 
\plotone{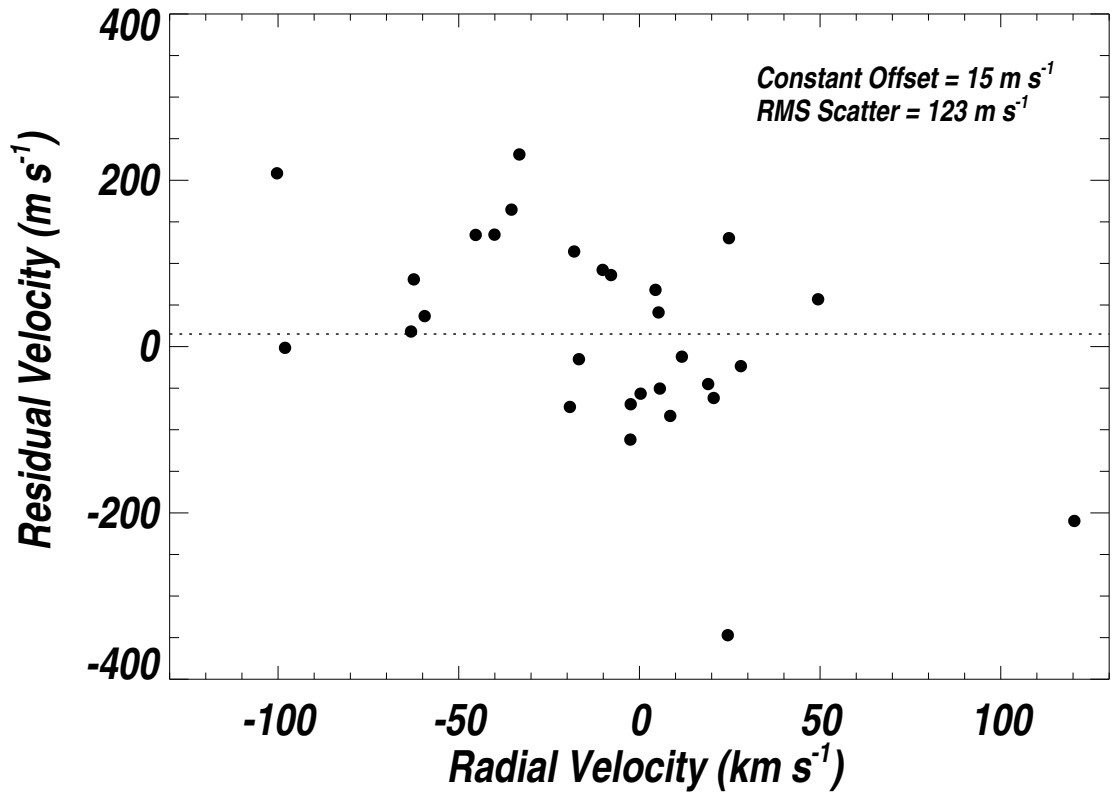} 
\caption{Standard--star velocities \citep{Stef99}  
minus present velocities for all 29 FGK stars in common.
The differences reveal that the present velocities 
are lower than those of Stefanik et al. by 15 \mse,  
and exhibit an rms scatter of 123 \mse. Thus the 
present velocities and those of \citet{Stef99}
differ in zero--point by $\sim$15 \mse.} 
\label{stef_res} 
\end{figure}

\begin{figure} 
\plotone{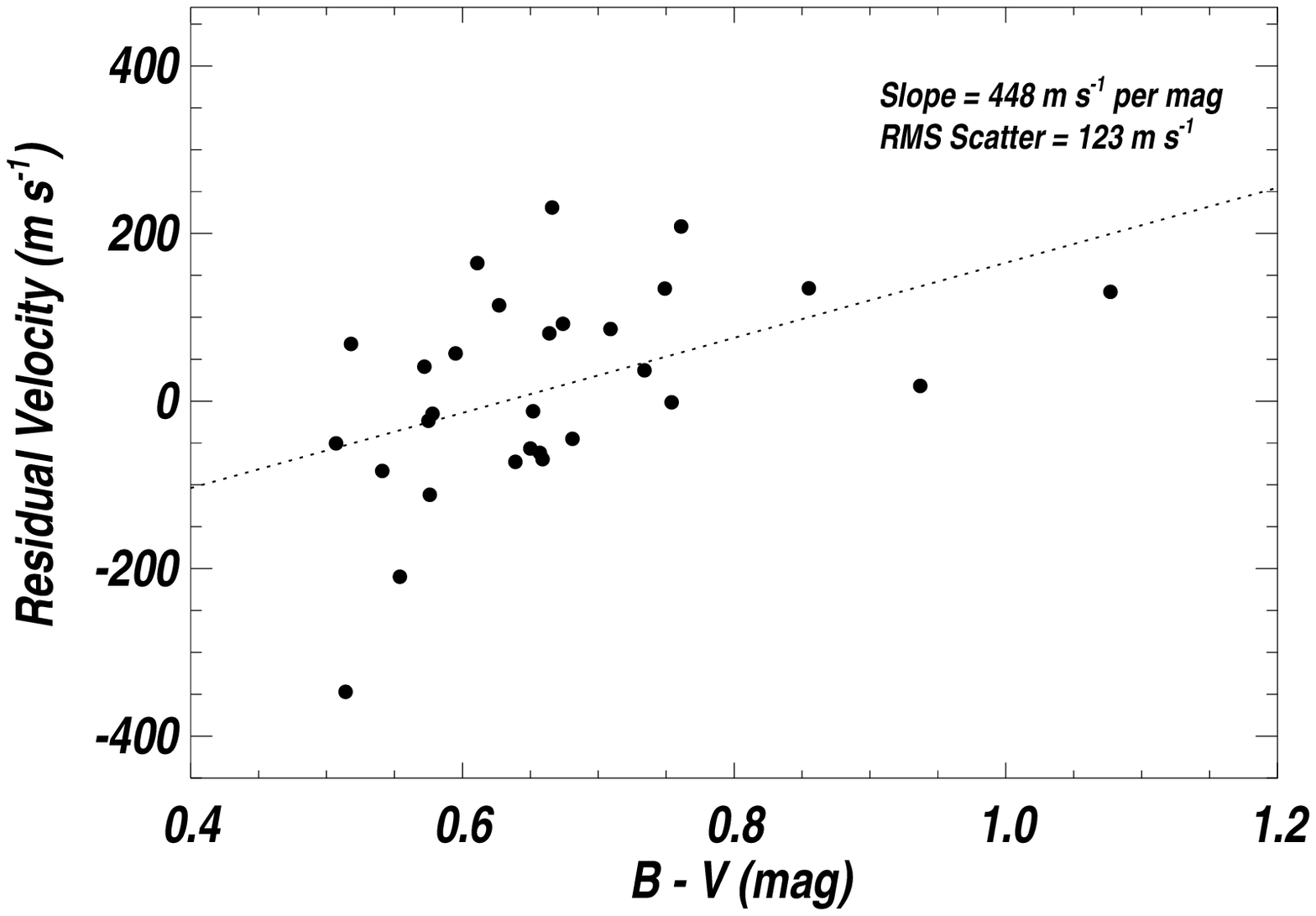} 
\caption{Standard--star velocities \citep{Stef99} minus the present velocities
as a function of \bv.  A dependence is apparent, suggesting systematic 
errors in at least one set of velocities.  The slope is $+448 \pm 111$ \ms 
per mag.  The RMS scatter is 123 \ms before fitting and 109 \ms after
fitting a line to the data.} 
\label{stef_bv} 
\end{figure}

Comparing our velocities for 29 common stars with those reported by 
\citet{Stef99}, with a correction of +136 \ms added to their native
velocities in Table 1 and 2 (Stefanik private communication), we
obtain $<V_{\rm CfA}-V_{\rm Present}>$ = 15 \ms which is only marginally different
from zero, with an RMS scatter of
123 \ms as seen in Figure \ref{stef_res}.  The differences exhibit
a significant \bv \ dependence as seen in Figure \ref{stef_bv}.  The 
slope of the linear trend is $+448 \pm 111$ \ms per mag with an RMS scatter
around the fit of 109 \mse.  This color dependence is likely caused
by some spectral type mismatch in one or all of the radial--velocity scales.
For solar type stars, we are in good agreement with the CfA velocities,
with no offset as seen in Figure \ref{stef_bv}.  This is likely due to
the CfA velocity zero--point being set by observations of
minor planets as was also done for the present velocities.
 
Interestingly, the sign of the slope in Figure \ref{stef_bv} is opposite
of that in Figure \ref{cor_bv} between the CORAVEL and present velocities.  
Since synthetic spectra were used to derive both the CfA and ELODIE velocities, 
which were used as the reference system in \citet{Udry99a},
\citep{Stef99,BQM96,Gul99} this seems to give credence to
the notion that using synthetic spectra does not entirely solve the problem of
spectral dependent systematic errors.

\subsection{M stars} 
 
\begin{figure} 
\plotone{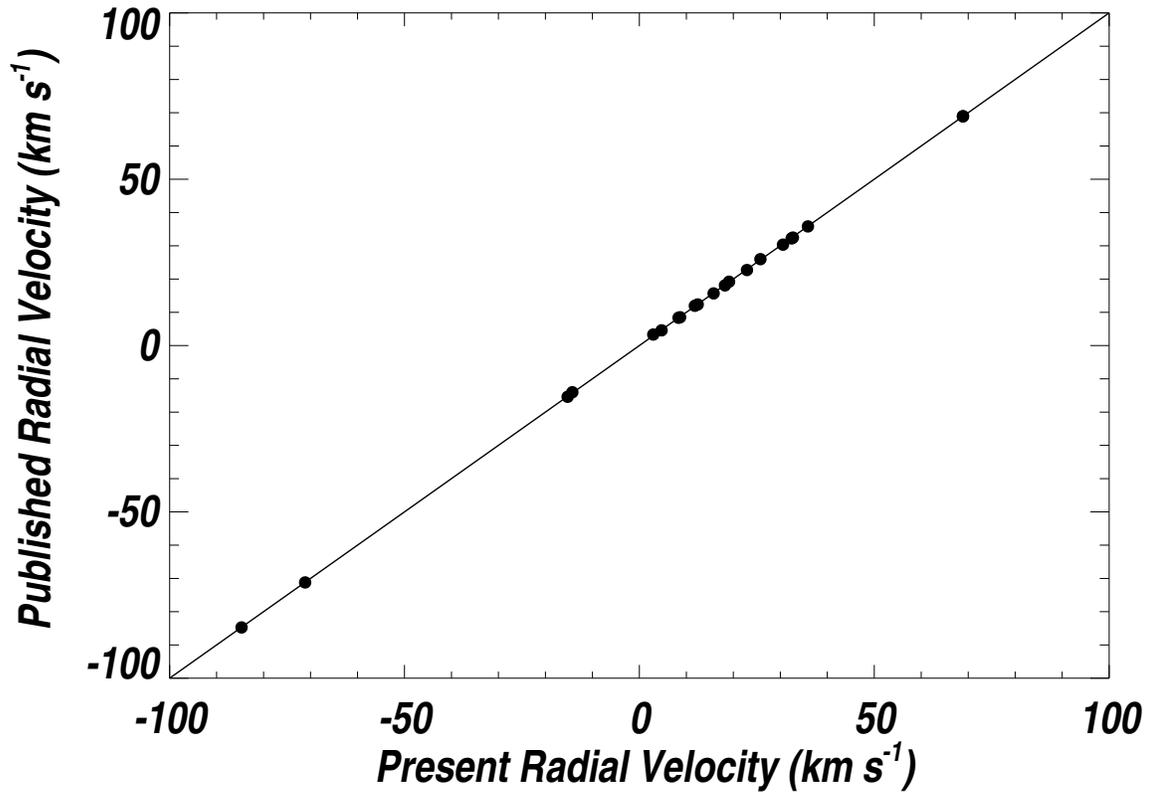} 
\caption{Velocities of standard stars (Marcy et al. 1987) vs. present velocities 
for M dwarfs.  The velocities agree well with no apparent nonlinear dependence.} 
\label{glw_lin} 
\end{figure}

\begin{figure} 
\plotone{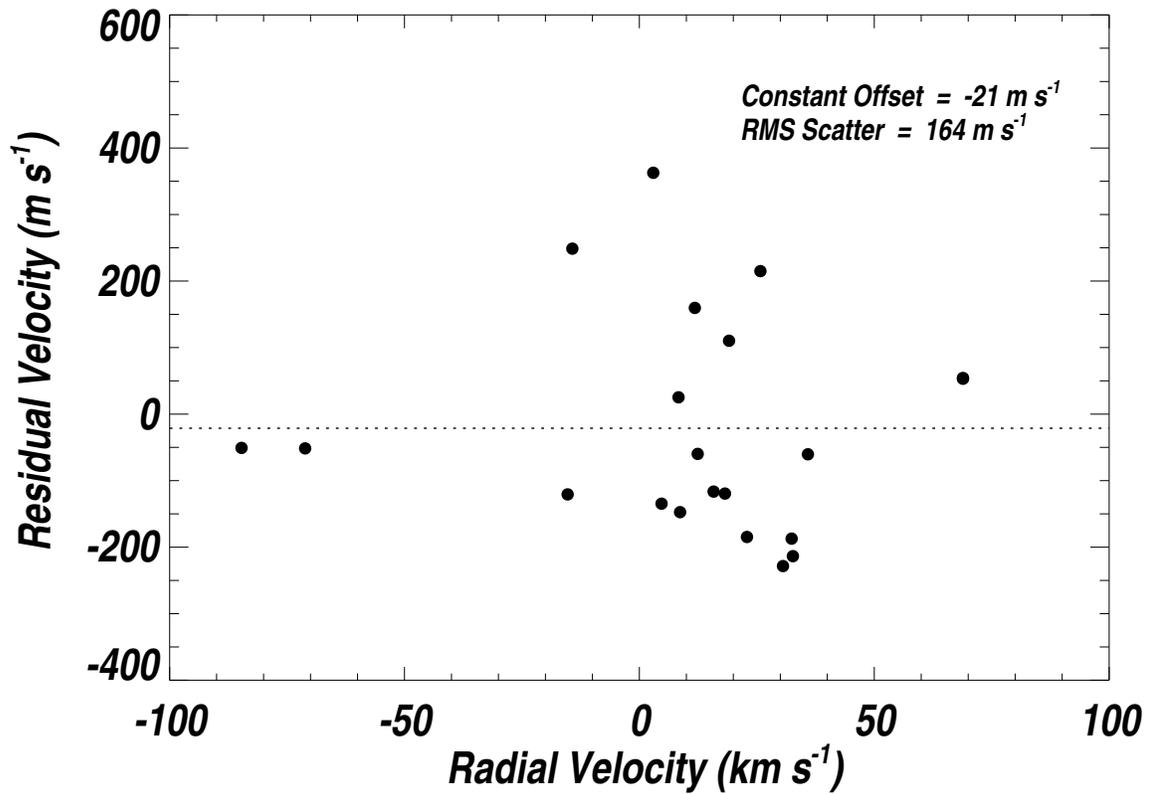} 
\caption{Difference between standard stars velocities \citep{Mar87} and
present velocities for all 21 M stars in common (as in Figure 7).
There is an rms scatter of 164 \mse, and a  
constant offset of -21 \ms which is not statistically significant. 
Thus our velocities for the stars having \bv $>$ 1.3 reside 
on the velocity scale set by \citet{Mar87}.  No \bv \ dependence is seen
in the residuals.} 
\label{glw_res} 
\end{figure}

The radial--velocity standard stars listed by \citet{Udry99a} do not 
include any M dwarfs, and we do not have any M dwarfs in common with 
\citet{Stef99} or the older CORAVEL standard stars \citep{Udry99b}. 
Therefore, we compared our present velocities for M dwarfs to those 
given in \citet{Mar87}.  The results of this comparison, for 21 stars 
in common, are shown in Figures \ref{glw_lin} and \ref{glw_res} and 
yield $<V_{MLW} - V_{Present}>$ = $-21$ \ms with an rms scatter of 164 
\mse.  The offset is obviously very low since the M star reference  
spectrum was created using the velocities quoted by \citet{Mar87}. 
Since the average internal error for the velocities in \citet{Mar87} is 
$\sim$200 \mse, most of the scatter in the differences is due to them. 
No \bv \ dependence is seen in the residuals. It is therefore difficult 
to ascertain the uncertainty in our present velocities for M dwarfs and 
similarly difficult to ascertain a zero--point error (whatever that 
would mean).  But a conservative estimate of the errors would be the
$\sim$0.4 \kms uncertainty of the \citet{Mar87} velocities which were
used in setting the velocity zero--point.
%Since the \citet{Mar87} velocities have published
%errors of 0.4 \kms a conservative zero--point error might be $\sim$0.4 \kmse.
 
Since the observed offset (21 \mse) for the M dwarfs is within a 
factor of two of the internal scatter of $\sim$10 \mse, we have not 
applied any correction to the radial velocities for these stars. 
There is some concern that our two sets of stars, the F, G, and K 
stars and the M stars, might not be on the same velocity zero-point. 
A comparison between them is difficult due to the significant spectral 
type mismatch errors.

\section{Uncertainty Estimates}

Due to the several systematic errors affecting radial velocities on the
order of 0.1 \kms it is difficult to asertain the true uncertainties
of the velocities reported here.  Normally two different methods are
used to estimate uncertainties: 1) Standard deviation of the mean
(i.e. the internal RMS scatter of points), and 2) comparison with
published values.  We have done both here.  On average the standard
deviation of the mean, due to the internal scatter of points, for the velocities
in Table 1 is $\sim$10 \mse.  We also compared our velocities with the
best known published standard star velocities of \citet{Udry99a}.  This
comparison, as shown in the previous section, gave an RMS scatter of the
differences of 35 \mse.  Therefore, using the traditional methods of
estimating uncertainties our velocites are accurate to $\sim$35 \mse.

In this case the traditional methods fail due to the astrophysical
sources of errors that affect all spectroscopic measurements of radial
velocity.  The values of these errors are not known adequately,
otherwise we would have corrected for them.  It is likely that these
systematic errors were also not taken into account by \citet{Udry99a}
or \citet{Stef99}.  This means that a comparison between their data
and ours will not yield the true uncertainty of our velocities or
theirs.

What the comparison does show is the precision of velocities within
a given spectral type.  The velocities of all stars of a given spectral
type will have a nearly constant offset from their true kinematic velocities,
because the systematic errors are  dependent on spectral type.
Relative to that constant offset the velocities in that spectral type
are very precise.  The comparison with \citet{Udry99a} shows our
precision to be no worse than 0.035 \kms (see Fig. 3).

This high precision within a spectral type can be effectively used
to look for moving groups.  Moving groups have velocity dispersions
of typicaly $\sim$0.5 \kms such as the Pleiades group \citep{Jon70}.  The present
velocities are precise enough within a spectral type to judge whether a
star belongs to the moving group or not.

Even though the exact values of the systematic errors are not known we
can estimate our true uncertainties.  There are three major systematic
errors in our velocities, namely, convective blueshift, gravitational
redshift and spectral type mismatch.  These errors change
systematically with spectral type.  Since our velocity zero--point was
set here using the day sky and the minor planet Vesta, which have well
known radial velocities due to Solar System dynamics, the systematic
errors were forced to zero for solar--type stars.  The errors will
rise as the spectral type departs from solar--type.  Because the
zero--point calibration was used for all FGK--type stars there will be
differential errors due to convective blueshift and gravitational
redshift that increase with departure from solar--type.  Similarly,
the spectral type mismatch errors only occur for non solar--type
stars, since the NSO solar spectrum was used for the reference, and is
therefore unaffected by the velocity zero--point calibration.

The approximate values of the systematic errors are as follows.
According to \citet{Dra99} the convective blueshift is approximately
--1000 \ms for F5V (\bv = 0.4), --400 \ms for G2V (solar-type, \bv =
0.64), and --200 \ms for K0V (\bv = 0.9).
The effective velocity error caused by gravitational redshift can be computed
from, $V_{\rm grav} = GM/Rc$ \citep{DGL99}.  We find,
with masses and radii given by \citet{All00}, that the redshift is
+680 \ms for F5V, +636 \ms for G2V, and +590 \ms for K0V.  The sum of
the two effects shows that the overall systematic error is
approximately --320 \ms for F5V, +236 \ms for G2V, and +390 \ms for
K0V.  Due to our zero--point calibration these errors are here forced
to zero for solar--type stars.  {\bf Therefore,  we expect that our
velocities are low by $\sim$556 \ms for F5V stars, true for
solar--type stars, and high by $\sim$154 \ms for K0V stars.}

With these estimates of the systematic errors in hand we can compute a correction
for the velocities as a function of \bv.  We fit a line to the first two points
(F5V and G2V) and to the second two points (G2V and K0V) to obtain two linear interpolations.
The first equation, Eqn (\ref{early}), is for main--sequence
stars earlier than solar--type (\bv $<$ 0.64), and the second equation, Eqn (\ref{late}),
is for main--sequence stars later than solar--type (\bv $>$ 0.64), not including M type stars.
Equation (\ref{late}) may be used for stars with 1.3 $>$ \bv $>$ 0.9, but 
represents an extrapolation
of the points given above and should be used with caution.  Since most of our main--sequence
stars have \bv $<$ 1.1 this shouldn't cause too many problems.  These equations do not
account for spectral type mismatch errors because we do not know enough about these effects
yet for our velocities.

\begin{equation}
\mbox{RV}_{\rm kin} = \mbox{RV}_{\rm spec} - 2.317 \times (\bv) + 1.483~ \kms \hspace{52pt} \mbox{\bv $<$ 0.64}
\label{early}
\end{equation} 

\begin{equation}
\mbox{RV}_{\rm kin} = \mbox{RV}_{\rm spec} - 0.642 \times (\bv) + 0.411~ \kms \hspace{20pt} \mbox{0.64 $<$ \bv $<$ 1.3}
\label{late}
\end{equation} 

These corrections are valid only for main--sequence stars
for which the NSO solar spectrum was used as the reference.  The velocities for
the M type stars, with \bv $>$ 1.3 and for which the M composite spectrum was
used as the reference, have a zero--point uncertainty of $\sim$0.4 \kms and therefore
a correction for gravitational redshift or convective blueshift is not warranted
at this time.  However, we expect the velocities of the M stars to be very
precise due to their low RMS velocity scatter and their small spectral
type range which minimizes the systematic errors.  We expect them to also be precise to
0.03 \kms, just as the FGK main--sequence stars.  If the ``true'' radial velocity
of only one of them were known then they all could be corrected for their zero--point
error and be very accurate.

We expect that the corrections given in equations (\ref{early}) and
(\ref{late}) account for convective blueshift and gravitational
redshift to within $\sim$0.1 \kmse.  The errors due to spectral type
mismatch are likely to be $\sim$0.1 \kms \citep{Grif00}.  From the
errors expressed in equations (\ref{early}) and (\ref{late}), an
additional $\sim$0.1 \kms for the spectral type mismatch error, along
with the distribution of spectral types of our survey stars, the
typical uncertainty of the uncorrected radial velocities in Tables 1
and 2 is $\sim$0.3 \kmse.  For the few subgiants, the errors are
somewhat larger, but not easily estimated without models of
subphotospheric convection in such stars.

%With the correction we presume that the radial velocities will be within $\sim$0.1 \kms of the
%``true'' kinematic radial velocities.  Of course, this does not account for any spectral type
%mismatch errors.
%These errors, if they exist, might be on the order of 0.1 \kms \citep{Grif00}.  If we
%add these errors in quadrature the uncertainty of the corrected barycentric radial velocites
%presented here is $\sim$0.15 \kmse.  The uncorrected barycentric radial velocities have
%typical uncertainties of $\sim$0.3 \kmse.

We cannot at this time give a correction for the velocities of our 45
giants.  Giants have a gravitational redshift on the order of 0.1 \kms
which is much smaller than that for dwarfs due to the large radius of
giants, $\sim$15 $R_{\odot}$ \citep{All00}.  The convective blueshift
is not known for giants and hinders us from giving a rough velocity
correction.  Hopefully, more in-depth future studies of the sources of
error discussed here, such as recent work by Pourbaix et al. (2002),
will allow for more accurate corrections of radial velocities and
eventually yield ``true'' radial velocities.  Until that time these
corrections may be used for the present velocities.

\section{Final Radial Velocities and Description of Tables} 
 
The barycentric radial velocities for all 889 stars are reported in 
Tables 1 and 2.  The 782 stars that exhibit an RMS velocity scatter less than 
100 \ms are reported in Table 1.  Primary and alternate star names are 
given in the first two columns, and the stellar spectrum used as the 
template (either NSO or M dwarf composite) is listed in column three. 
The mean time of the observations is given in column four under 
$<$JD$>$ to establish the characteristic epoch for the velocity 
measurement.  The span of observations in days is given in column 
five, and the mean barycentric radial velocity of all observations for a 
star is in column six.  Stars with only one observation were put in 
Table 1 even though their RMS scatter is not defined.  They can be 
distinguished by $\Delta$T = 0. 
 
The 107 stars with an RMS velocity scatter greater than 100 \ms are 
reported in Table 2.  Primary and alternate star names are given in 
the first two columns.  The stellar template spectrum (either NSO or M 
dwarf) is given in column three.  The Julian Date (JD) of one specific 
observation is given in column four.  The barycentric radial velocity of 
that one observation is given in column five.  The mean date of all 
observations and span of observations are given in columns six and 
seven.  The RMS scatter of the velocities of all observations is given 
in column eight, and the number of observations in column nine.  The 
last column is for comments where stars with companion orbits or 
linear trends are noted.  ``L'' indicates that velocities vary 
linearly with time (see Table 4), ``CO'' indicates that a companion 
and its orbit were found, ``C'' indicates that a published companion exists
but an orbit is not given here,  
%exists in a long--period orbit, but we could not determine that orbit, 
and ''A'' indicates that the star is chromospherically active based on 
the emission reversal seen at the CaII H\&K lines in our spectra. 
 
The orbital parameters for 15 stars with companions are reported in 
Table 3.  The orbital period $P$, velocity semi--amplitude $K$, 
eccentricity $e$, longitude of periastron $\omega$, and time of periastron $T_0$ are given as 
well as the primary mass $M_1$, minimum secondary mass $M_{\rm 2,min}$, minimum semi--major axis $a_{\rm min}$,
and the mass function, $f(M)$.  References are also given to other 
sources which have information on these stars, their companions and 
orbital parameters. The data for stars with linear trends are given in 
Table 4.  The slope of the radial velocity curve and the number of 
observations is given for 30 stars.

\section{Orbits of Binaries} 
 
We found 107 stars in this program that exhibit an RMS velocity 
scatter greater than 100 \mse.  We attempted to fit these
with a Keplerian orbit.  We found 29 stars that yield good Keplerian orbital 
fits to their relative radial velocities.  Many of these binaries have been 
previously published from our velocities and therefore will not be 
duplicated here.  Several papers contain these previously reported 
single--line spectroscopic binaries, namely Marcy et al. (1999), 
Butler et al. (2000), Vogt et al. (2002), Fischer et al. (2002), 
Cumming et al. (1999), Marcy et al. (2001b). 
 
For 15 stars, our velocities provide unpublished orbital solutions or
reveal unknown companions.  These orbits are listed in Table 3 which
gives the usual orbital parameters for single--line spectroscopic
binaries.  Plots of the velocities and the associated Keplerian fits
are shown in Figures 9--23.  The mass for the primary star for each
system was estimated from the catalog by \citet{Pri99} or by using \bv
\ and \citet{All00}.  From the orbital parameters and primary masses,
we determined the minimum companion masses (\mmine), mass functions, $f(M)$,
and minimum semimajor axes, $a_{\rm min}$, which are also listed in Table 3.  The values of
\mmin range from 42 M$_{\rm Jup}$ to 545 M$_{\rm Jup}$, and therefore
the companions are all candidate brown dwarfs or H--burning stellar
companions.  The uncertainties in the orbital parameters were found by
a Monte Carlo technique in which Gaussian velocity noise was added to
the best--fit theoretical velocity curve at the times of observation.
We ran 50 trials for each case, with orbital parameters being
rederived for each trial.  The standard deviations of the resulting
orbital parameters were taken as the uncertainties.
For some stars the Monte Carlo method underestimated the uncertainties
in the orbital parameters.  For these stars a more conservative
estimate of the uncertainties was made by fitting many different Keplerian
orbits to the observed relative radial velocities, without any modeled
Gaussian noise, and then looking at the scatter of the parameters
produced by the best orbital fits.

The stars HD4747, HD65430 and GJ84 have large orbital uncertainties
due to incomplete phase coverage.  There were insufficient velocity
measurements to constrain the orbit of HD18445 even though it is known
to have a companion \citep{Hal00}.  Thus it isn't listed in Table 3.
The eccentricity for HD140913 was fixed to $e=0.54$ \citep{Lam89} 
in performing the Keplerian fit,  since not enough points were available to constrain the
eccentricity.  Only the other four parameters were fit for this star.
The Keplerian fit for HD208776 is particularly poor as
insufficient velocities are available to constrain the orbital period
to better than a factor of two.  Eleven of the spectroscopic binary
stars appear to be newly discovered here: HD4747, HD7483, HD30339,
HD34101, HD39587, HD65430, HD174457, HD208776, GJ84, GJ595, HIP52940.
Velocities are available upon request of G.M.  The companions all have
\mmin in the sub--stellar range or low mass stellar range, and thus
offer interesting targets for studies with adaptive optics or
interferometry.

\section{Conclusion}

We have provided barycentric radial velocities with an internal precision
of 0.03 \kms for 889 stars.
The error estimates stem both from the internal 
errors found from the spectral chunks within each spectrum and from 
the comparison with accurate velocities on the CORAVEL scale 
\citep{Udry99a}.  The radial velocities of the F, G and K dwarfs 
reside on a velocity zero--point defined by the observations of the 
Sun, using the day sky and Vesta as proxies.  Our velocity scale 
differs by only 0.053 \kms from that of \citet{Udry99a} and 0.015 \kms
from that of \citet{Stef99}, thus adding 
confidence to the zero points of all three sets of velocities.  The radial 
velocities of the M dwarfs reside on the velocity system defined by 
\citet{Mar87} and have not been further corrected, nor is such a 
correction known to be necessary.  These M dwarf velocities are 
probably accurate to within 200 \mse.    

The Doppler shifts reported
here have such high accuracy that gravitational redshift and
convective blueshift impose comparable (or greater) wavelength
shifts.  These effects were somewhat removed from our velocity measurements
by using the Sun for the velocity zero--point.  We expect therefore that for G2V 
stars the present velocities represent their ``true'' kinematic velocities within 0.03 \kmse.
However, for stars departing from solar--type the sum of the two astrophysical effects
will produce systematic errors dependent on spectral type.  From F to K
type dwarfs this variation will be as large as $\sim$0.3 \kms and will cause
our velocities to be low for F type stars and high for K type stars.  Using
estimates for these effects we give rough velocity corrections
in equations (\ref{early}) and (\ref{late}).  We presume that these corrections
bring the velocities within $\sim$0.15 \kms of their ``true'' kinematic
values.

These precise radial velocities can be used to complement
future highly precise proper motion measurement, such as those
projected to be obtained by the GAIA mission of the ESA.  
The radial velocities and proper motions will give the
three components of space motion of stars.
 %which can be used for several purposes.
These precise space motions may be useful for discerning
the membership of moving groups, since young moving groups
have velocity dispersions of $\sim$0.5 \kms \citep{Jon70}. 

The 782 stars listed in Table 1 exhibited a velocity scatter of less than 100
\ms during 4 years.  These stars apparently  exhibit relatively stable
velocities on time scales of a decade  and represent candidates for
radial velocity standard stars.  However, their integrity as velocity
standard stars requires   future observations to verify their
stability.    We expect that some of these 782 ``stable'' stars may
reveal slow drifts in  radial velocity on time scales longer than 10
yr.  

The accuracy of the  present velocities offers an opportunity to
detect such slow drifts by future measurements made with comparable
accuracy.  Such velocity variations may prove useful in identifying
unseen companions  at large orbital distances, i.e., over 10 AU.    We found that 107
stars exhibited velocity variations of over 100 \ms  (RMS).  For these
stars, we have provided the RMS velocity, and also  either an orbital
solution or a description of the linear trends.  We  intend these
measurements to provide dynamical constraints on the  nature of the
companions.   
\acknowledgements

We acknowledge support by NSF grant AST-9988358 (to SSV), NSF grant 
AST-9988087 and travel support from the Carnegie Institution of 
Washington (to RPB), NASA grant NAG5-8299 and NSF grant AST95-20443 
(to GWM), and by Sun Microsystems.  We thank the NASA and UC Telescope 
assignment committees for allocations of Keck telescope time, and we 
thank the University of California for generous allocations of 
telescope time at Lick Observatory.  This research has made use of the 
Simbad database, operated at CDS, Strasbourg, France.  We would like 
to thank A. Reines, C. McCarthy, E.J. Anderson, and S. White 
for all of their help and inspiration.

\clearpage 
 
%-figure 1 of avg. obs. err vs. b-v 
 
%-figure 2 of linear trend (CORAVEL-Present) 
 
%-figure 3 of residuals (CORAVEL-Present) 
 
%-figure 4 of residuals vs. b-v (CORAVEL-Present) 

%-figure 5 of residuals (Stefanik-Present) 

%-figure 6 of residuals vs. b-v (Stefanik-Present) 

%-figure 7 of linear trend (GLW-Present) 
 
%-figure 8 of residuals (GLW-Present) 

%-figure 9 of binary HD 4747
\begin{figure}
\centerline{\scalebox{.75}{\rotatebox{90}{\includegraphics{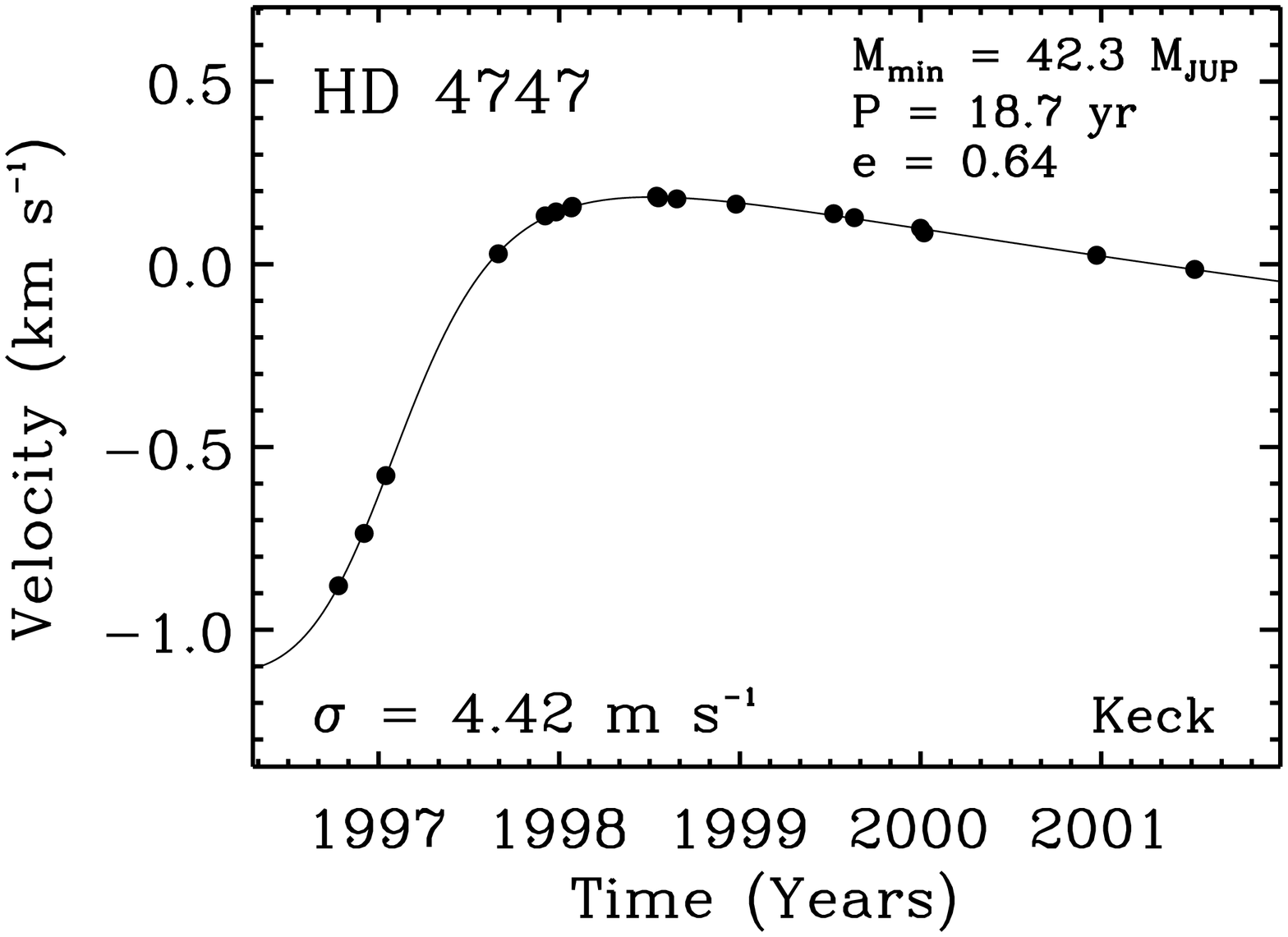}}}}
%\plotone{f9.ps}
\caption{Doppler velocities for HD 4747 (G8/K0 V).
The solid line is a Keplerian orbital fit with a
period of 18.7 yr, a semiamplitude of 0.65 \kmse,
and an eccentricity of 0.64, yielding a minimum
(\mmine) of 42.3 \mjup for the companion.  The
RMS of the Keplerian fit is 4.42 \mse.}
\label{4747plot}
\end{figure} 

%-figure 10 of binary HD 7483
\begin{figure}
\centerline{\scalebox{.75}{\rotatebox{90}{\includegraphics{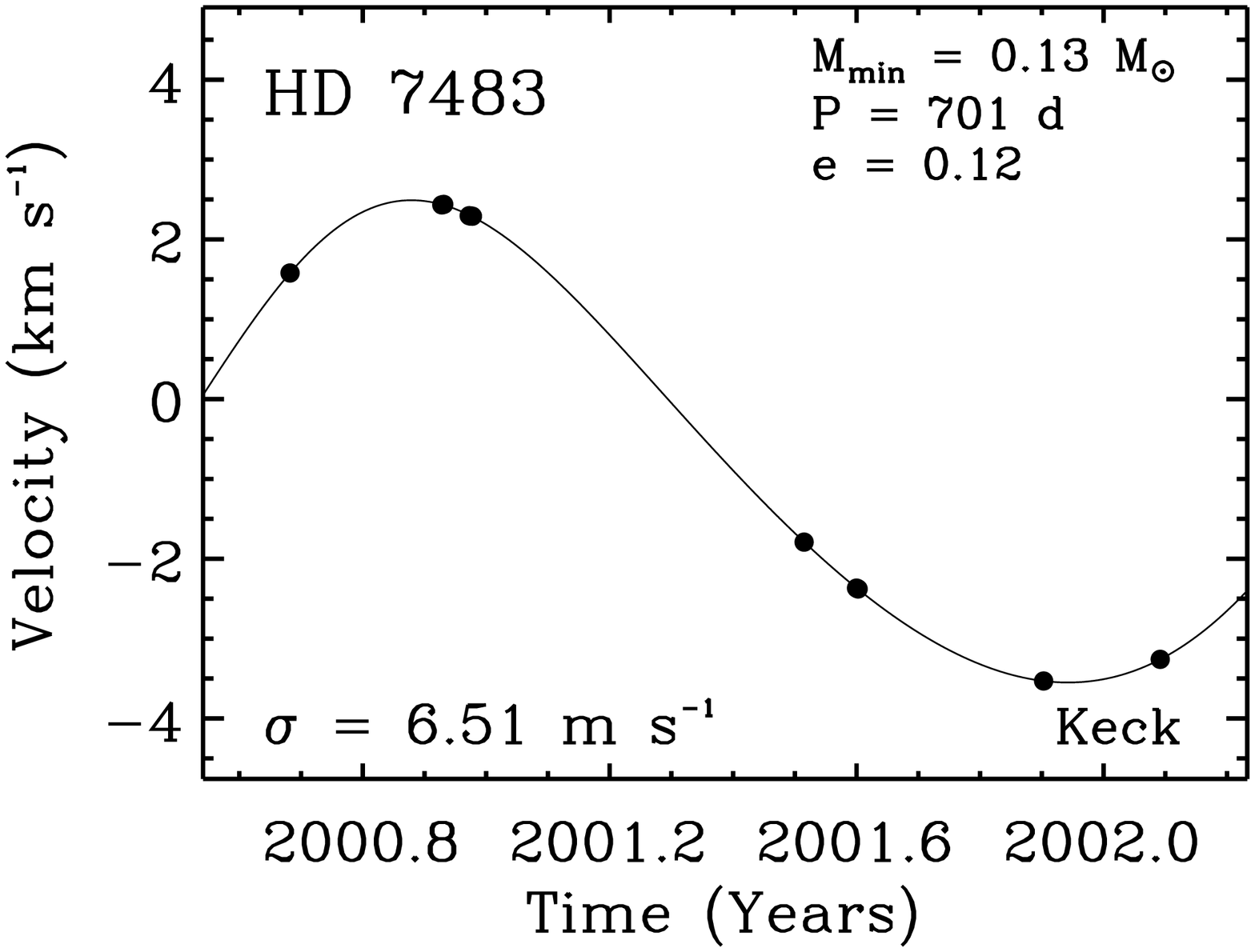}}}}
%\plotone{f10.ps}
\caption{Doppler velocities for HD 7483 (G5 V).
The solid line is a Keplerian orbital fit with a
period of 701 d, a semiamplitude of 3.02 \kmse,
and an eccentricity of 0.12, yielding a minimum
(\mmine) of 0.13 \msun for the companion.  The
RMS of the Keplerian fit is 6.51 \mse.}
\label{7483plot}
\end{figure} 

%-figure 11 of binary HD 30339
\begin{figure}
\centerline{\scalebox{.75}{\rotatebox{90}{\includegraphics{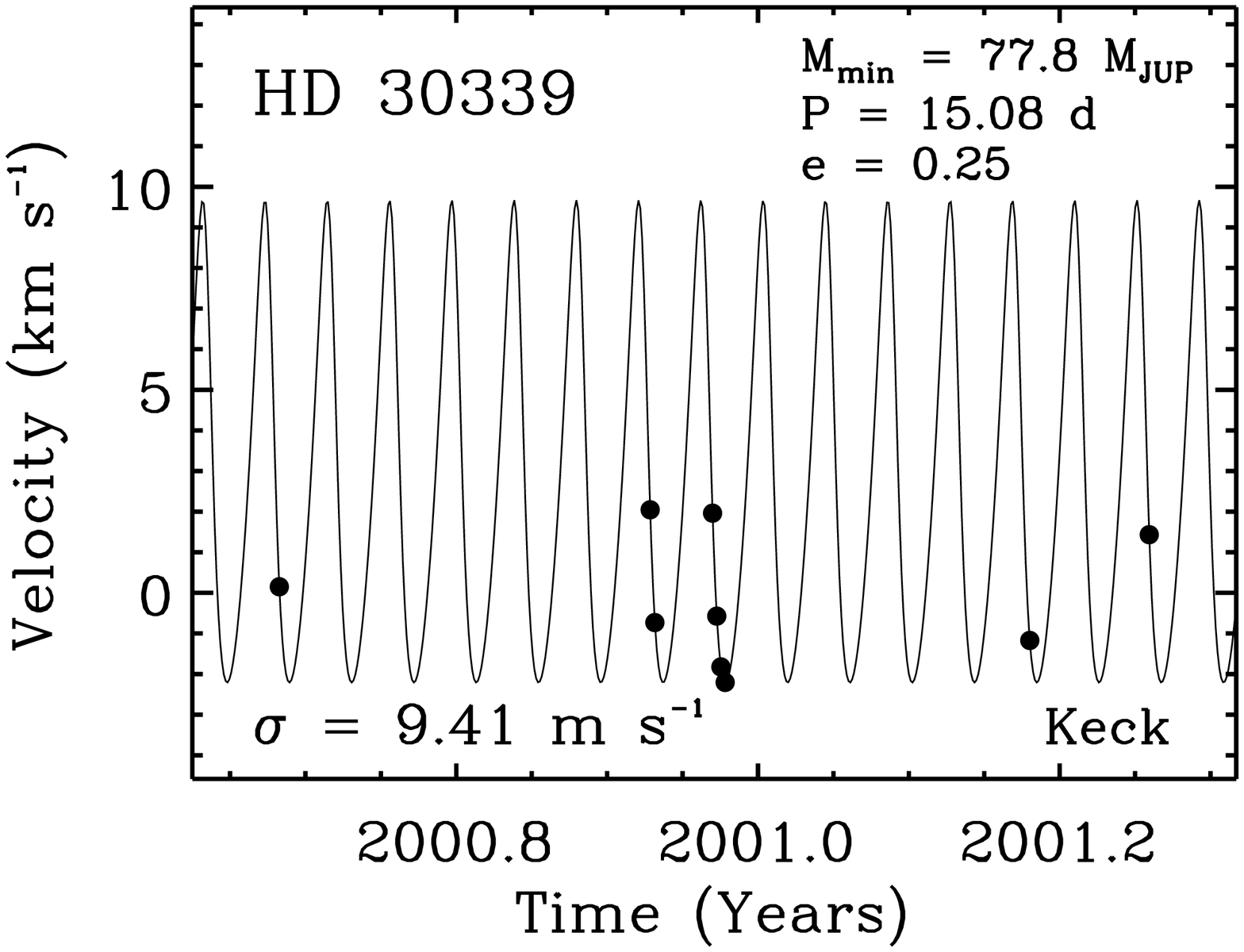}}}}
%\plotone{f11.ps}
\caption{Doppler velocities for HD 30339 (F8 V).
The solid line is a Keplerian orbital fit with a
period of 15.08 d, a semiamplitude of 5.94 \kmse,
and an eccentricity of 0.25, yielding a minimum
(\mmine) of 77.8 \mjup for the companion.  The
RMS of the Keplerian fit is 9.41 \mse.}
\label{30339plot}
\end{figure} 

%-figure 12 of binary HD 34101
\begin{figure}
\centerline{\scalebox{.75}{\rotatebox{90}{\includegraphics{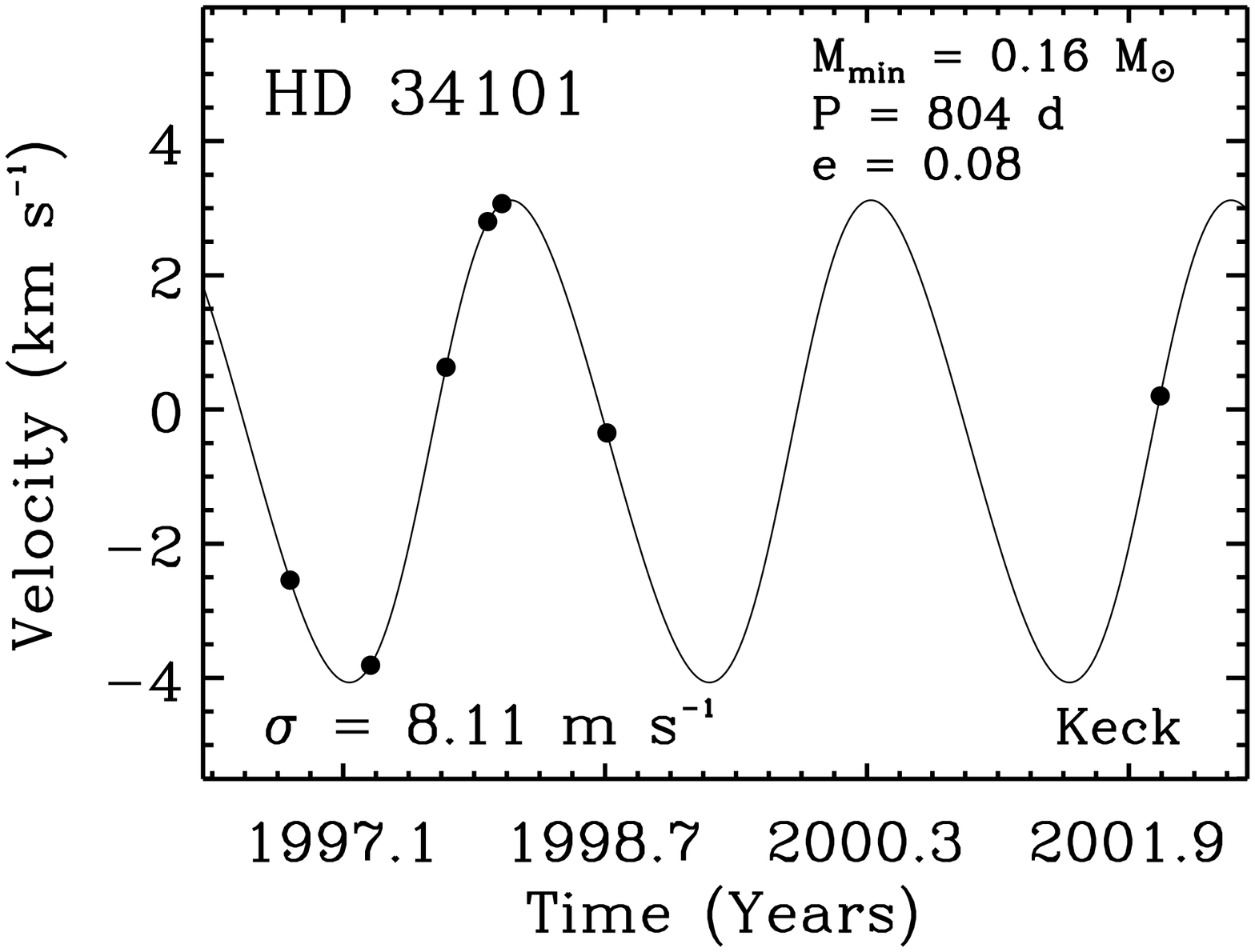}}}}
%\plotone{f12.ps}
\caption{Doppler velocities for HD 34101 (G8 V).
The solid line is a Keplerian orbital fit with a
period of 804 d, a semiamplitude of 3.76 \kmse,
and an eccentricity of 0.08, yielding a minimum
(\mmine) of 0.16 \msun for the companion.
The RMS of the Keplerian fit is 8.11 \mse.}
\label{34101plot}
\end{figure} 

%-figure 13 of binary HD 39587 (2047 lick)
\begin{figure}
\centerline{\scalebox{.75}{\rotatebox{90}{\includegraphics{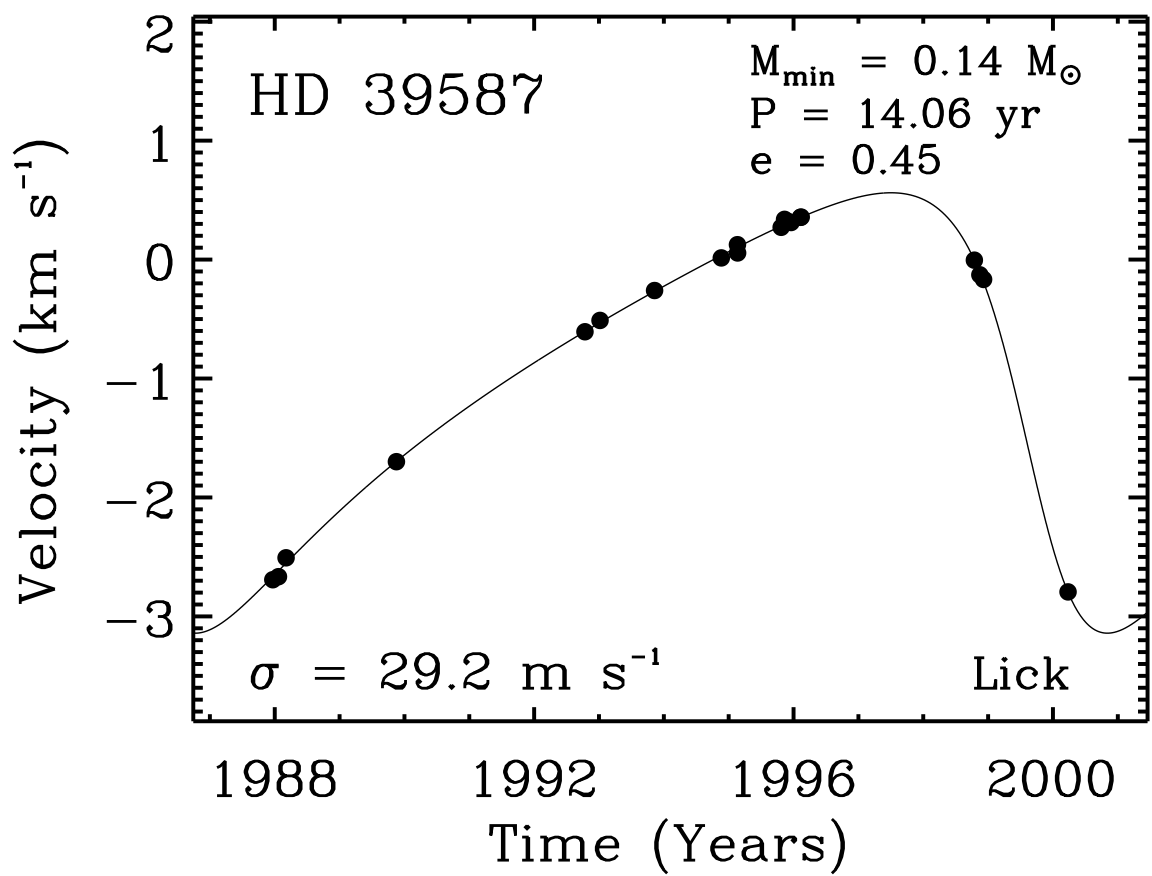}}}}
%\plotone{f13.ps}
\caption{Doppler velocities for HD 39587 (G0 V).
The solid line is a Keplerian orbital fit with a
period of 14.06 yr, a semiamplitude of 1.85 \kmse,
and an eccentricity of 0.45, yielding a minimum
(\mmine) of 0.14 \msun for the companion.  The
RMS of the Keplerian fit is 29.2 \mse.}
\label{39587plot}
\end{figure} 
\clearpage

%-figure 14 of binary HD 65430
\begin{figure}
\centerline{\scalebox{.75}{\rotatebox{90}{\includegraphics{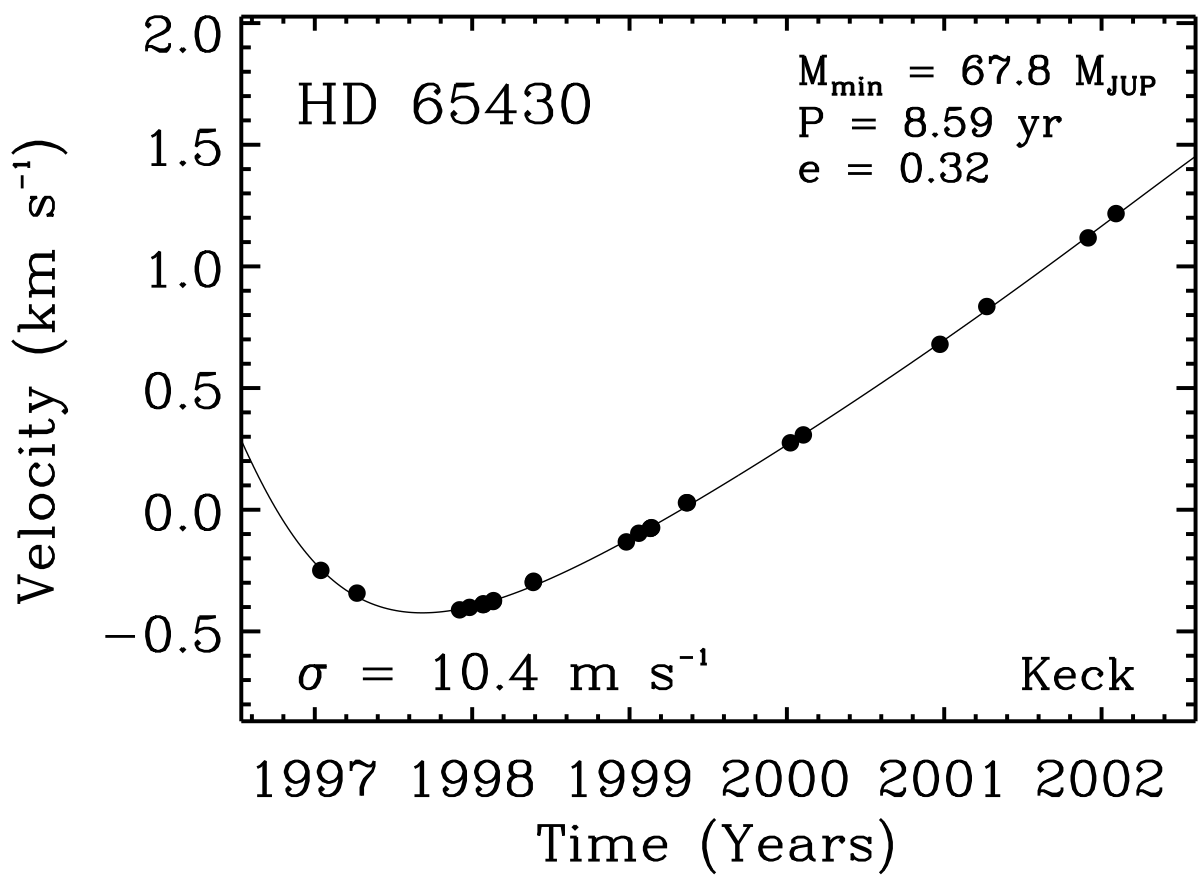}}}}
%\plotone{f14.ps}
\caption{Doppler velocities for HD 65430 (K0 V).
The solid line is a Keplerian orbital fit with a
period of 8.59 yr, a semiamplitude of 1.11 \kmse,
and an eccentricity of 0.32, yielding a minimum
(\mmine) of 67.8 \mjup for the companion.  The
RMS of the Keplerian fit is 10.4 \mse.}
\label{65430plot}
\end{figure} 
\clearpage

%-figure 15 of binary HD 122742 (5273 lick)
\begin{figure}
\centerline{\scalebox{.75}{\rotatebox{90}{\includegraphics{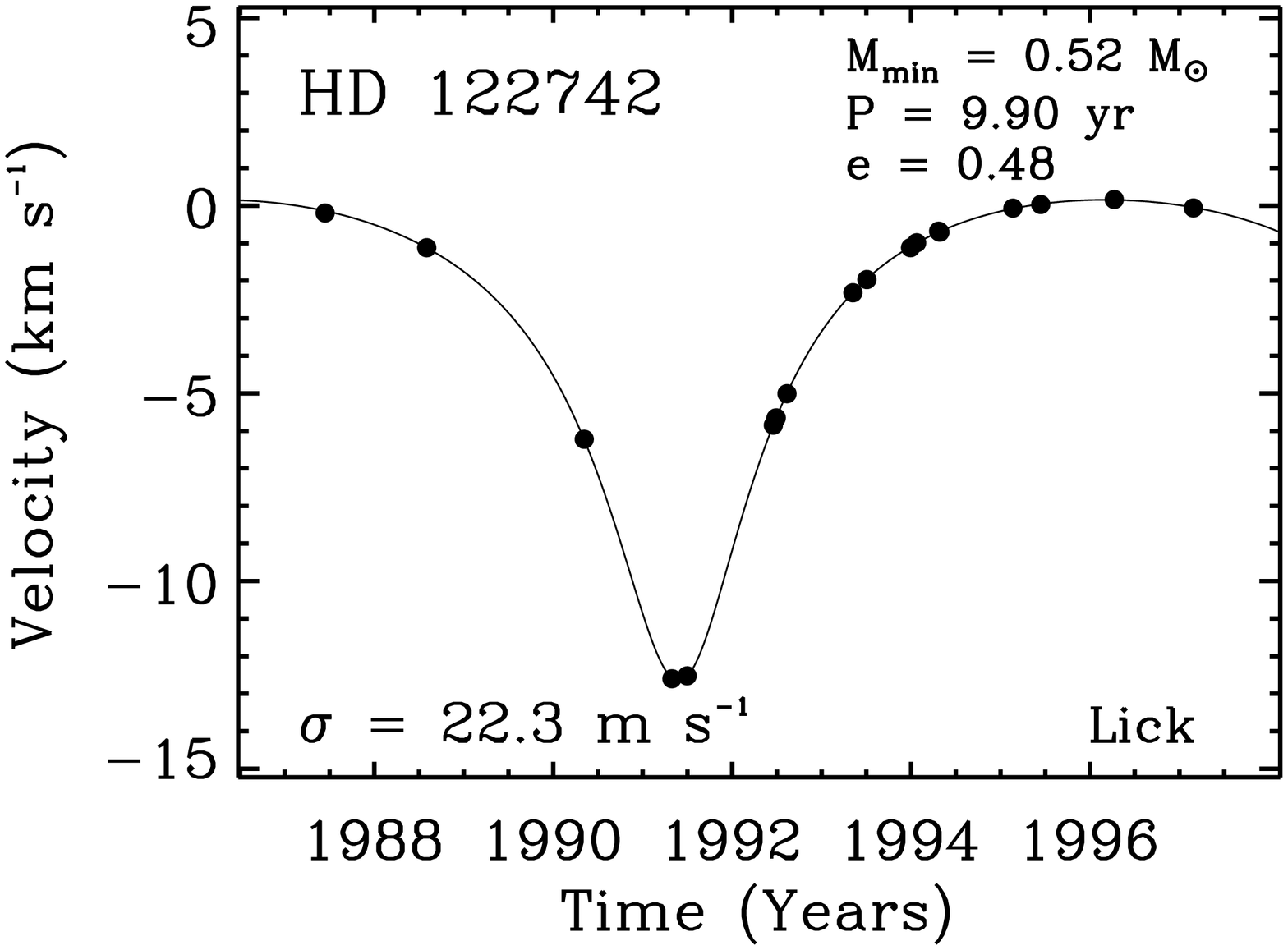}}}}
%\plotone{f15.ps}
\caption{Doppler velocities for HD 122742 (G8 V).
The solid line is a Keplerian orbital fit with a
period of 9.90 yr, a semiamplitude of 6.41 \kmse,
and an eccentricity of 0.48, yielding a minimum
(\mmine) of 0.52 \msun for the companion.  The
RMS of the Keplerian fit is 22.3 \mse.}
\label{122742plot}
\end{figure} 
\clearpage

%-figure 16 of binary HD 131511 (5553 lick)
\begin{figure}
\centerline{\scalebox{.75}{\rotatebox{90}{\includegraphics{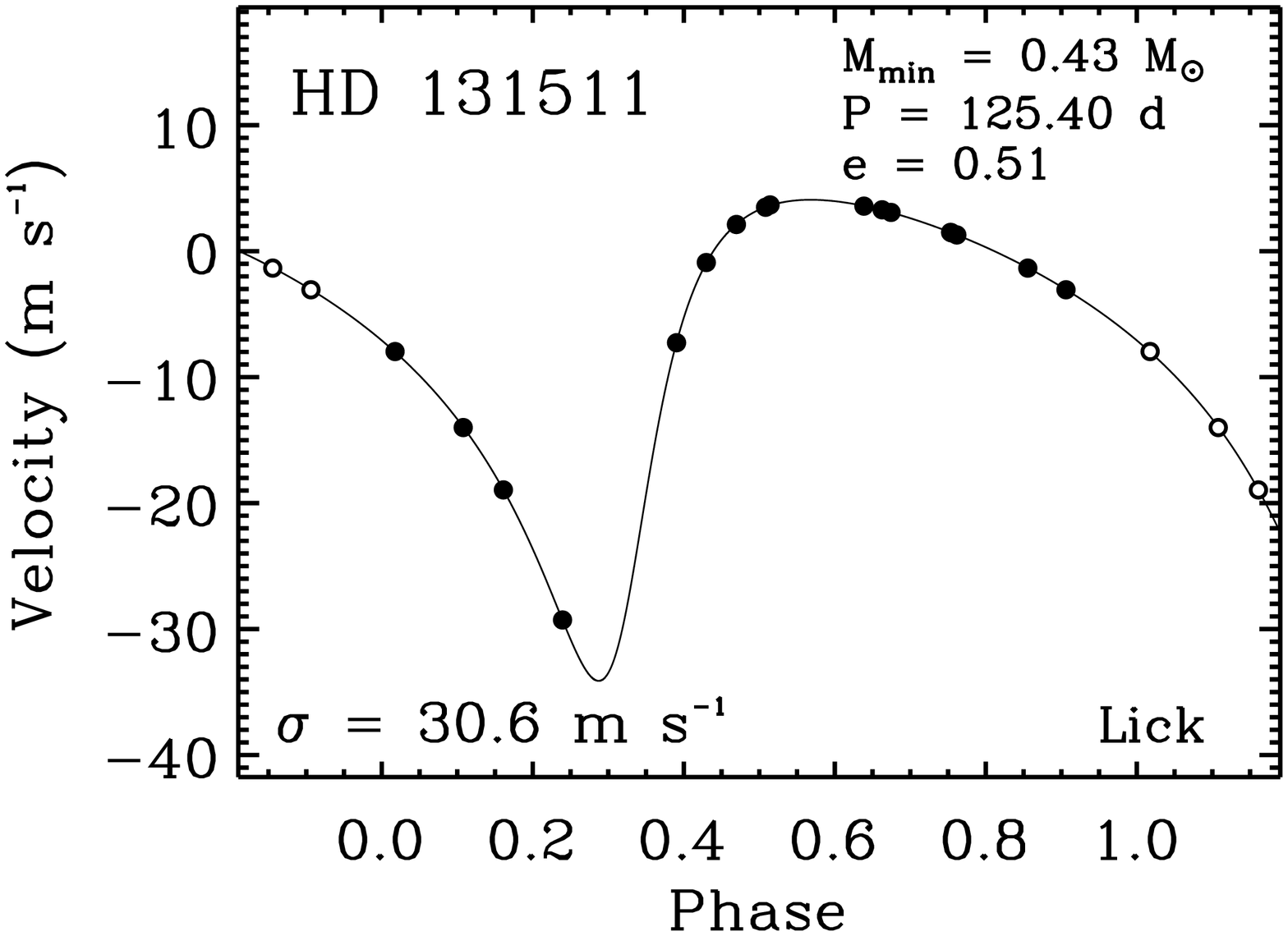}}}}
%\plotone{f16.ps}
\caption{Doppler velocities for HD 131511 (K2 V).
The solid line is a Keplerian orbital fit with a
period of 125.40 d, a semiamplitude of 19.10 \kmse,
and an eccentricity of 0.51, yielding a minimum
(\mmine) of 0.43 \msun for the companion.  The
RMS of the Keplerian fit is 30.6 \mse.}
\label{131511plot}
\end{figure} 
\clearpage

%-figure 17 of binary HD 140913
\begin{figure}
\centerline{\scalebox{.75}{\rotatebox{90}{\includegraphics{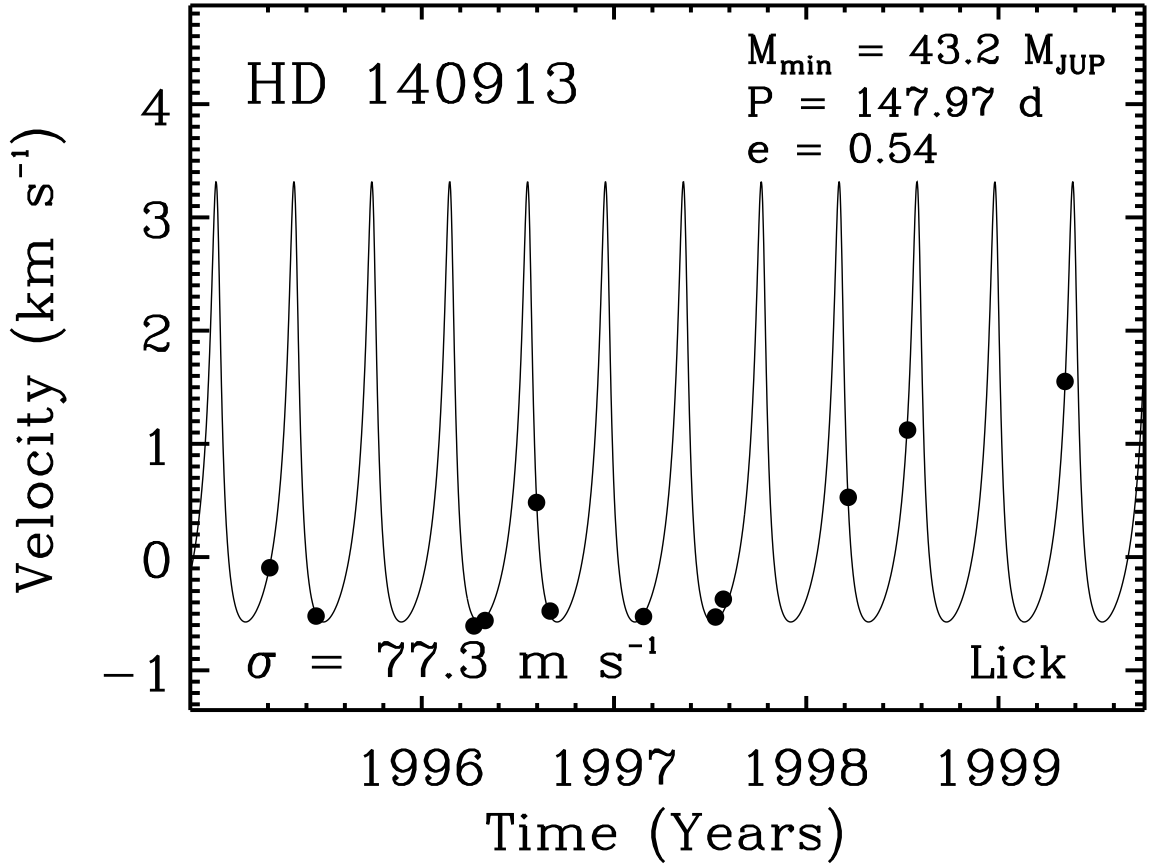}}}}
%\plotone{f17.ps}
\caption{Doppler velocities for HD 140913 (G0 V).
The solid line is a Keplerian orbital fit with a
period of 147.97 d, a semiamplitude of 1.94 \kmse,
yielding a minimum (\mmine) of 43.2 \mjup for the companion.
The eccentricity was fixed to $e=0.54$ given by Latham et al. (1989)
since not enough points were available to constrain the eccentricity.
The RMS of the Keplerian fit is 77.3 \mse.}
\label{140913plot}
\end{figure} 
\clearpage

%-figure 18 of binary HD 174457
\begin{figure}
\centerline{\scalebox{.75}{\rotatebox{90}{\includegraphics{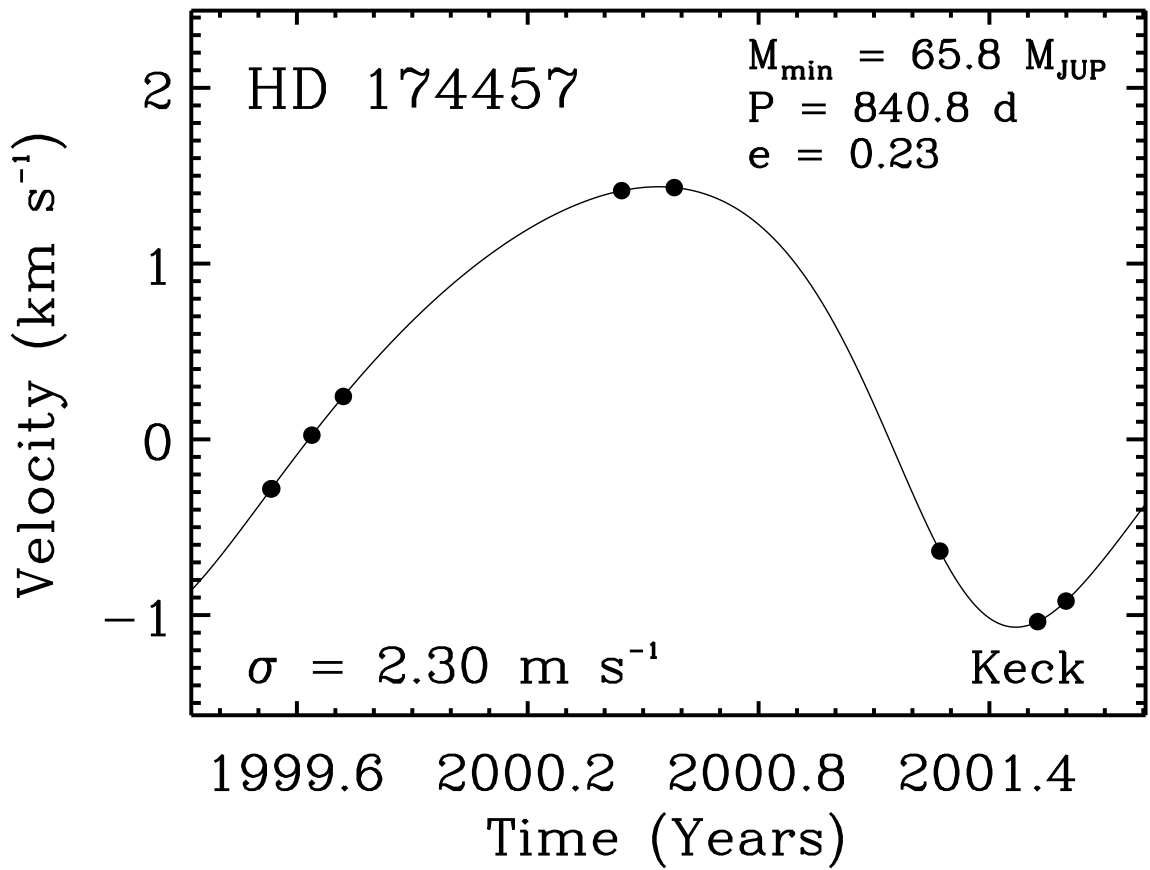}}}}
%\plotone{f18.ps}
\caption{Doppler velocities for HD 174457 (F8 V).
The solid line is a Keplerian orbital fit with a
period of 840.8 d, a semiamplitude of 1.25 \kmse,
and an eccentricity of 0.23, yielding a minimum
(\mmine) of 65.8 \mjup for the companion.  The
RMS of the Keplerian fit is 2.30 \mse.}
\label{174457plot}
\end{figure} 
\clearpage

%-figure 19 of binary HD 208776
\begin{figure}
\centerline{\scalebox{.75}{\rotatebox{90}{\includegraphics{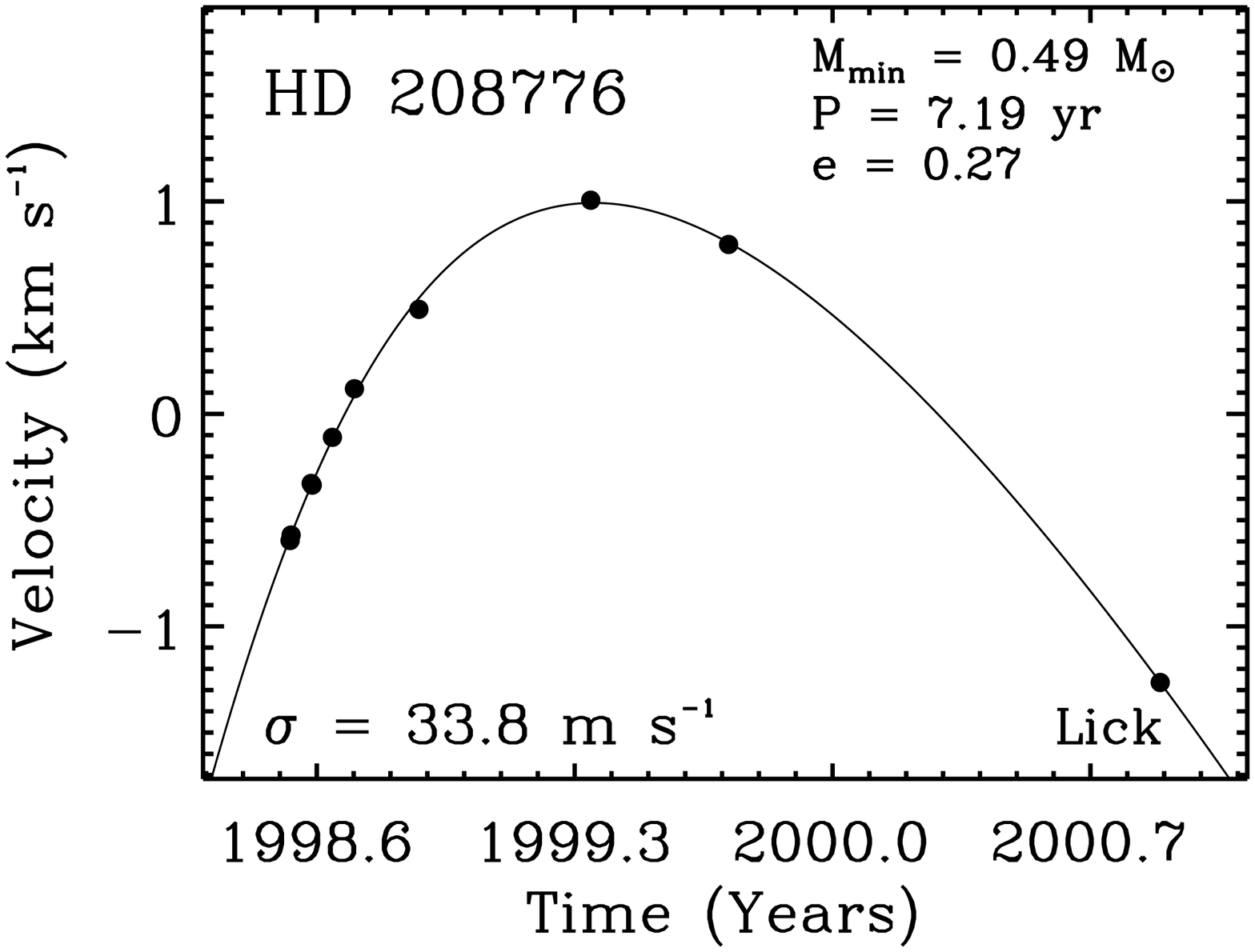}}}}
%\plotone{f19.ps}
\caption{Doppler velocities for HD 208776 (G0 V).
The solid line is a Keplerian orbital fit with a
period of 7.19 yr, a semiamplitude of 5.46 \kmse,
and an eccentricity of 0.27, yielding a minimum
(\mmine) of 0.49 \msun for the companion.  The
RMS of the Keplerian fit is 33.8 \mse.}
\label{208776plot}
\end{figure} 

\clearpage

%-figure 20 of binary GJ 84
\begin{figure}
\centerline{\scalebox{.75}{\rotatebox{90}{\includegraphics{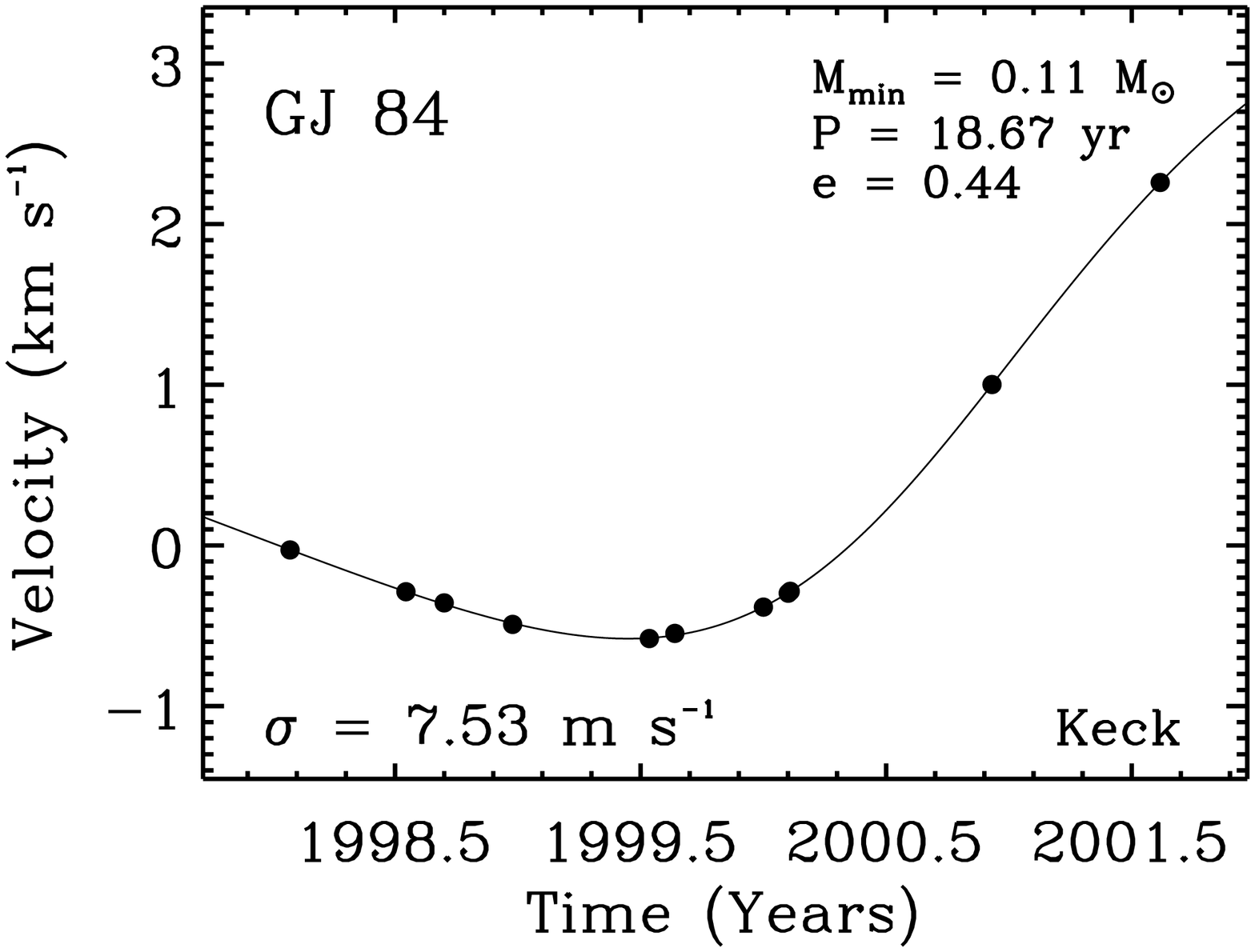}}}}
%\plotone{f20.ps}
\caption{Doppler velocities for GJ 84 (M3 V).
The solid line is a Keplerian orbital fit with a
period of 18.67 yr, a semiamplitude of 2.18 \kmse,
and an eccentricity of 0.44, yielding a minimum
(\mmine) of 0.11 \msun for the companion.  The
RMS of the Keplerian fit is 7.53 \mse.}
\label{gl84plot}
\end{figure} 

\clearpage

%-figure 21 of binary GJ 595 (hip76901 keck)
\begin{figure}
\centerline{\scalebox{.75}{\rotatebox{90}{\includegraphics{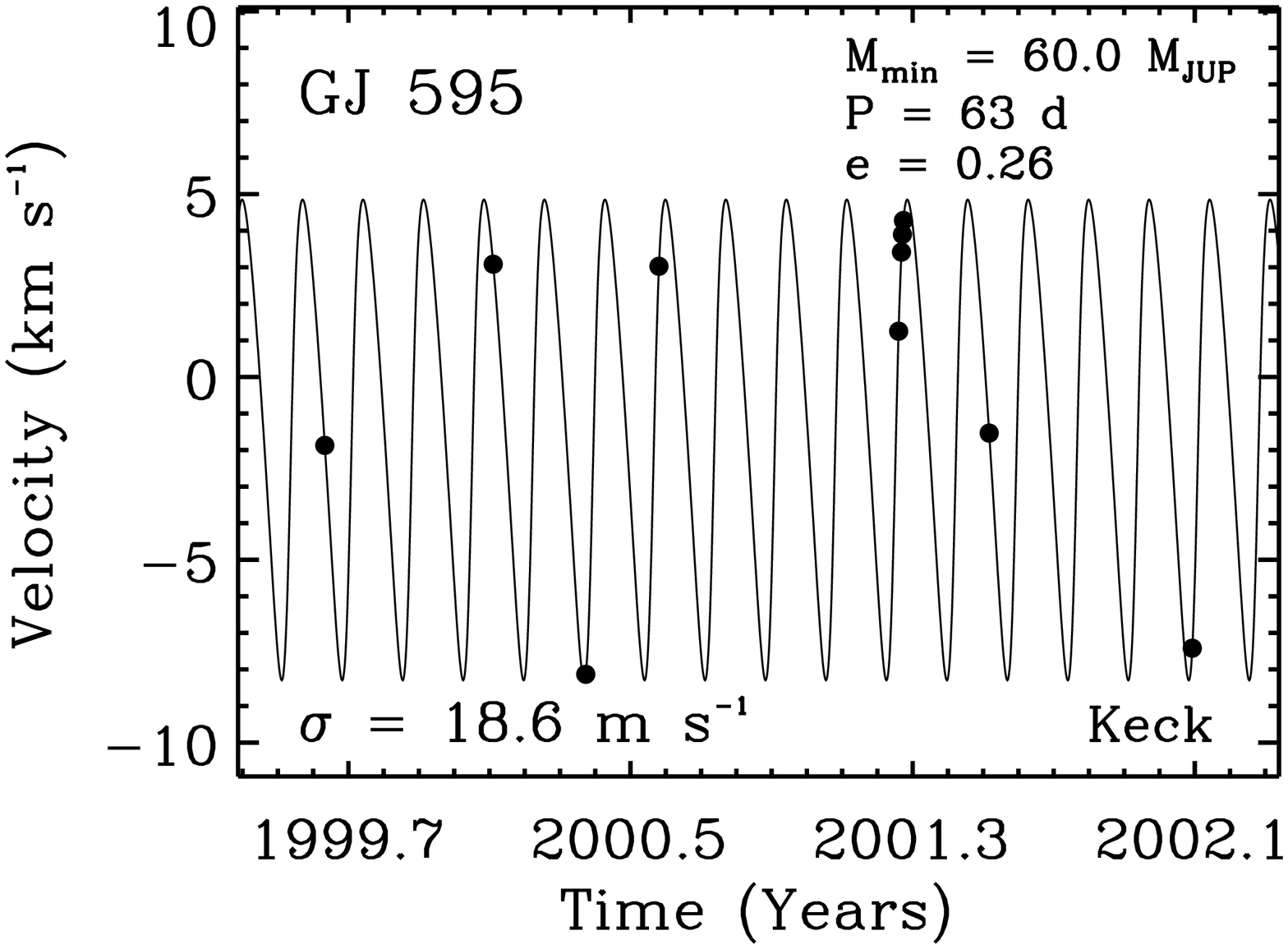}}}}
%\plotone{f21.ps}
\caption{Doppler velocities for GJ 595 (M3 V).
The solid line is a Keplerian orbital fit with a
period of 63 d, a semiamplitude of 6.57 \kmse,
and an eccentricity of 0.26, yielding a minimum
(\mmine) of 60.0 \mjup for the companion.  The
RMS of the Keplerian fit is 18.6 \mse.}
\label{gl595plot}
\end{figure} 

\clearpage

%-figure 22 of binary GJ 623
\begin{figure}
\centerline{\scalebox{.75}{\rotatebox{90}{\includegraphics{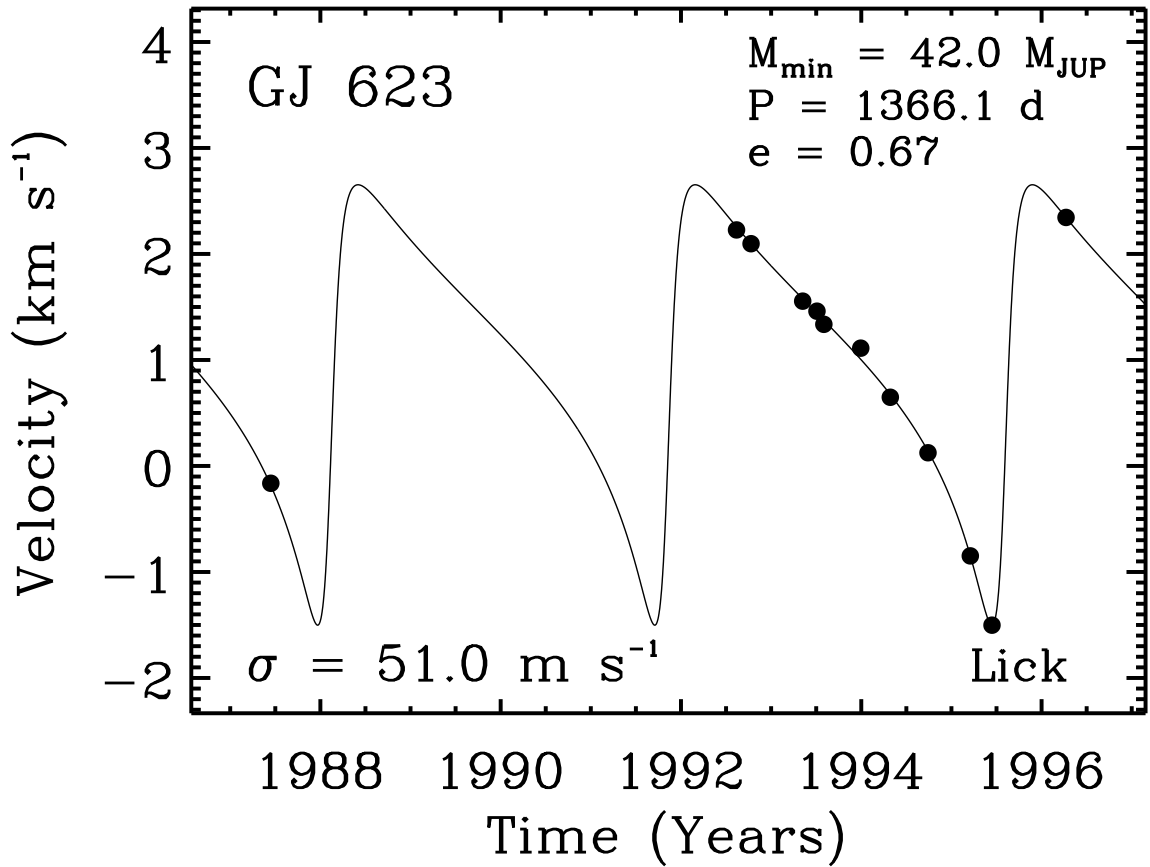}}}}
%\plotone{f22.ps}
\caption{Doppler velocities for GJ 623 (M3 V).
The solid line is a Keplerian orbital fit with a
period of 1366.1 d, a semiamplitude of 2.08 \kmse,
and an eccentricity of 0.67, yielding a minimum
(\mmine) of 42.0 \mjup for the companion.  The
RMS of the Keplerian fit is 51.0 \mse.}
\label{gl623plot}
\end{figure}
 
\clearpage

%-figure 23 of binary HIP 52940 (hip52942b keck)
\begin{figure}
\centerline{\scalebox{.75}{\rotatebox{90}{\includegraphics{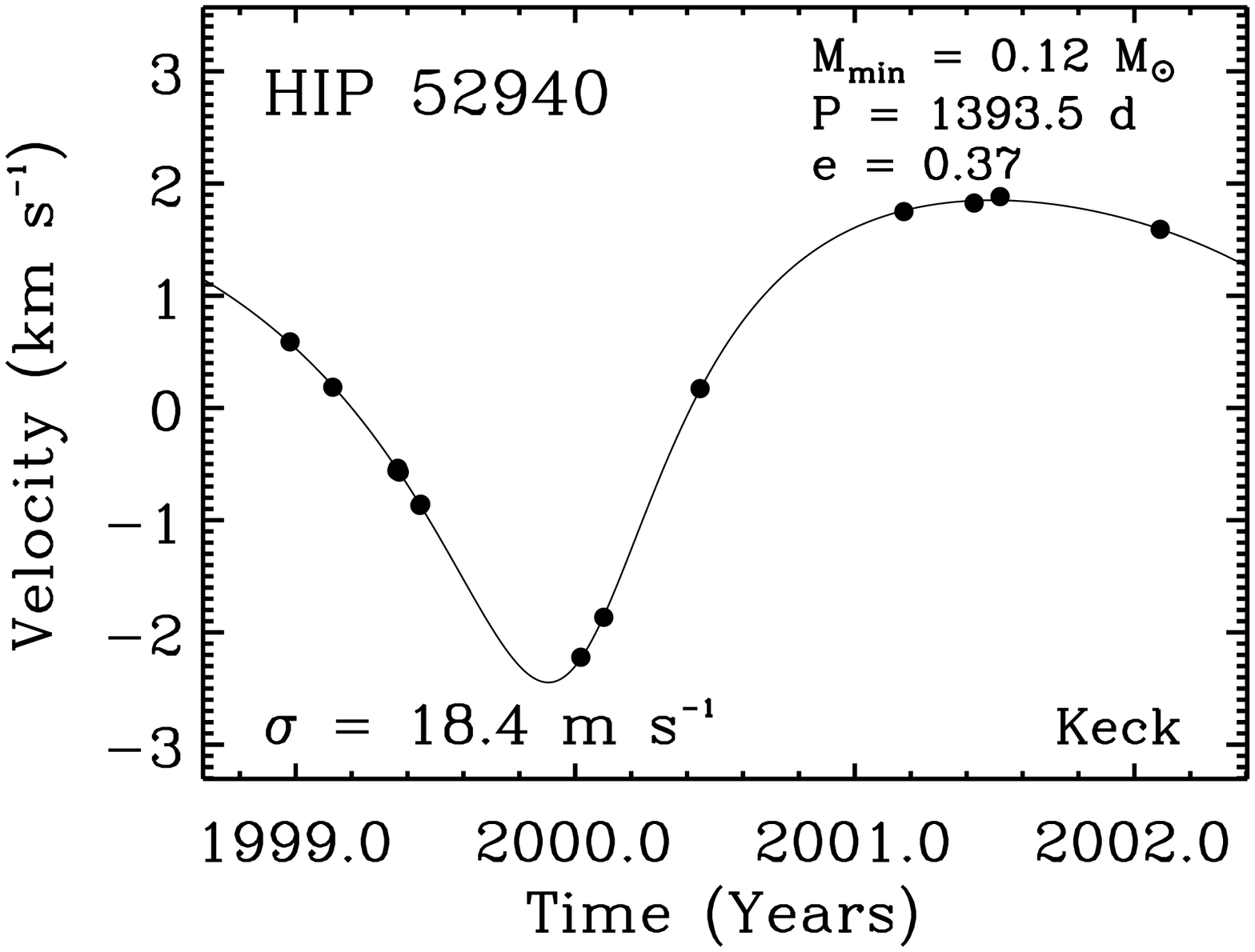}}}}
%\plotone{f23.ps}
\caption{Doppler velocities for HIP 52940 (F8 V).
The solid line is a Keplerian orbital fit with a
period of 1393.5 d, a semiamplitude of 2.15 \kmse,
and an eccentricity of 0.37, yielding a minimum
(\mmine) of 0.12 \msun for the companion.  The
RMS of the Keplerian fit is 18.4 \mse.}
\label{hip52940plot}
\end{figure} 

\clearpage

\begin{deluxetable}{llcrrr}
\tablenum{1}
\tablecaption{Radial Velocities of Stable Stars \tablenotemark{a}}
\tabletypesize{\small}
\label{rv1}
\tablewidth{0pt}
\tablehead{

\colhead{Primary Name} & \colhead{Alternate Name} & \colhead{Ref} & \colhead{$<$JD$>$} & \colhead{$\Delta$T} & \colhead{$<$RV$>$} \\
\colhead{} & \colhead{} & \colhead{(NSO/M)} & \colhead{($-$2,440,000)} & \colhead{(days)} & \colhead{(m s$^{-1}$)} \\
%%\colhead{Primary Name} & \colhead{Alternate Name} & \colhead{Ref} & \colhead{RV} \\
%%\colhead{} & \colhead{} & \colhead{(NSO/M)} & \colhead{(m s$^{-1}$)}

}
\startdata
\tableline
        \phm{A} HD 166  &      \phm{aa} HIP 544  &   NSO  &    $ 8,783 $ \phm{aa}  &         $ 2,662 $  &    $ -6,537 $ \phm{.}   \\
        \phm{A} HD 283  &      \phm{aa} HIP 616  &   NSO  &   $ 11,002 $ \phm{aa}  &         $ 1,185 $  &   $ -43,102 $ \phm{.}   \\
        \phm{A} HD 377  &      \phm{aa} HIP 682  &   NSO  &   $ 11,417 $ \phm{aa}  &           $ 208 $  &     $ 1,184 $ \phm{.}   \\
        \phm{A} HD 400  &      \phm{aa} HIP 699  &   NSO  &   $ 11,137 $ \phm{aa}  &           $ 507 $  &   $ -15,141 $ \phm{.}   \\
        \phm{A} HD 531  &     \phm{aa} BD +07 9  &   NSO  &   $ 11,419 $ \phm{aa}  &           $ 208 $  &    $ 13,460 $ \phm{.}   \\
      \phm{A} HD 1326A  &       \phm{aa} GJ 15A  &     M  &    $ 9,928 $ \phm{aa}  &         $ 4,656 $  &    $ 11,814 $ \phm{.}   \\
      \phm{A} HD 1326B  &       \phm{aa} GJ 15B  &     M  &   $ 11,294 $ \phm{aa}  &           $ 509 $  &    $ 10,976 $ \phm{.}   \\
       \phm{A} HD 1388  &     \phm{aa} HIP 1444  &   NSO  &   $ 11,016 $ \phm{aa}  &         $ 1,338 $  &    $ 28,498 $ \phm{.}   \\
\enddata
\tablenotetext{a}{ Stars with $\sigma_{\rm rms}$ $<$ 100 m s$^{-1}$.}
\end{deluxetable}

\eject

\begin{deluxetable}{llcrrrrrrc}
\tablenum{2}
\tablecaption{Radial Velocities:Stars with RMS $>$ 0.1 \kms \tablenotemark{a}}
\tabletypesize{\small}
\rotate
\label{rv2}
\tablewidth{0pt}
\tablehead{

\colhead{Primary} & \colhead{Alternate} & \colhead{Ref} & \colhead{JD} & \colhead{RV} & \colhead{$<$JD$>$} & \colhead{$\Delta T$} & 
          \colhead{$\sigma_{\rm rms}$} & \colhead{$N$} & \colhead{Com}\\
\colhead{Name} & \colhead{Name} & \colhead{(NSO/M)} & \colhead{($-$2,440,000)} & \colhead{(m s$^{-1}$)} & \colhead{($-$2,440,000)} & \colhead{(days)} &
\colhead{(m s$^{-1}$)} & \colhead{} & \colhead{}
%%\colhead{Primary Name} & \colhead{Alternate Name} & \colhead{Ref} & \colhead{JD} & \colhead{RV} & \colhead{$\Delta$RV} \\
%%\colhead{} & \colhead{} & \colhead{(NSO/M)} & \colhead{(-2,440,000)} & \colhead{(m s$^{-1}$)} & \colhead{(m s$^{-1}$)}
}
\startdata
\tableline

     \phm{a} HD 1854 &   \phm{a} SAO 192490 &    NSO &  $ 10984.112 $ \phn &    $ 24,327 $ \phn &     $ 10,714 $ \phm{aa.} &                  $ 119 $ &        $ 3,651 $ \phd &           $ 4 $ &              \\
     \phm{a} HD 3346 &     \phm{a} HIP 2900 &    NSO &  $ 10012.747 $ \phn &   $ -33,377 $ \phn &      $ 9,317 $ \phm{aa.} &                $ 2,220 $ &          $ 172 $ \phd &          $ 29 $ &              \\
     \phm{a} HD 3770 &     \phm{a} HIP 3169 &    NSO &  $ 11774.903 $ \phn &    $ -8,124 $ \phn &     $ 11,555 $ \phm{aa.} &                $ 1,132 $ &          $ 178 $ \phd &          $ 13 $ &          L   \\
     \phm{a} HD 3795 &     \phm{a} HIP 3185 &    NSO &  $ 11756.021 $ \phn &   $ -46,595 $ \phn &     $ 11,131 $ \phm{aa.} &                $ 1,535 $ &          $ 156 $ \phd &          $ 24 $ &          L   \\
     \phm{a} HD 4271 &     \phm{a} HIP 3540 &    NSO &  $ 11047.927 $ \phn &    $ 32,566 $ \phn &     $ 11,037 $ \phm{aa.} &                   $ 21 $ &        $ 7,799 $ \phd &           $ 2 $ &              \\
     \phm{a} HD 4747 &     \phm{a} HIP 3850 &    NSO &  $ 11756.051 $ \phn &    $ 10,399 $ \phn &     $ 11,092 $ \phm{aa.} &                $ 1,731 $ &          $ 315 $ \phd &          $ 21 $ &         CO   \\
     \phm{a} HD 7483 &     \phm{a} HIP 5881 &    NSO &  $ 12129.043 $ \phn &   $ -17,717 $ \phn &     $ 12,015 $ \phm{aa.} &                  $ 515 $ &        $ 2,602 $ \phd &          $ 11 $ &         CO   \\
     \phm{a} HD 8331 &     \phm{a} HIP 6442 &    NSO &  $ 11154.627 $ \phn &   $ -12,270 $ \phn &     $ 11,091 $ \phm{aa.} &                  $ 128 $ &        $ 2,349 $ \phd &           $ 2 $ &              \\
\enddata
\tablenotetext{a}{ Stars with $\sigma_{\rm rms}$ $\ge$ 100 m s$^{-1}$.}
\end{deluxetable}

\eject

\begin{deluxetable}{lrrrrrrrrrl}  
\tablenum{3}  
\tablecaption{Orbital Parameters}  
\tabletypesize{\small}  
\rotate  
\label{rv3}  
\tablewidth{0pt} \tablehead{ \colhead{Star} & \colhead{$P$} &  
\colhead{$K$} & \colhead{$e$} & \colhead{$\omega$} & \colhead{$T_0$} &  
\colhead{$M_1$} & \colhead{$M_{\rm 2,min}$} & \colhead{$a_{\rm min}$} & \colhead{$f(M)$} &
\colhead{Other} \\
\colhead{} & \colhead{(days)} & \colhead{(km s$^{-1}$)} & \colhead{} & \colhead{(deg)} &  
\colhead{(-2450000)} & \colhead{($M_{\odot}$)} & \colhead{($M_{\rm Jup}$)} &  
\colhead{(AU)} & \colhead{($M_{\odot}$)} & \colhead{References} \\ } \startdata  
\tableline   
             \phm{a} HD 4747 &       6832 (653) &     0.65 (0.1) &      0.64 (0.06) &          257 (5) &         453 (473) &   0.83  &  42.3\phm{a} &   6.7\phm{A} & 0.000087 &                \\
             \phm{a} HD 7483 &    701.42 (0.01) &    3.02 (0.01) &     0.12 (0.003) &        313 (0.5) &      1791.1 (0.7) &   0.92  &   136\phm{a} &   1.6\phm{A} & 0.0020   &                \\
            \phm{a} HD 30339 & 15.0778 (0.0003) &    5.94 (0.01) &     0.25 (0.001) &           43 (1) & 1881.199 (0.0003) &   1.10  &  77.8\phm{a} &  0.13\phm{A} & 0.00030  &                \\
            \phm{a} HD 34101 &    803.51 (0.03) &    3.59 (0.01) &     0.08 (0.001) &          275 (1) &           690 (2) &   0.87  &   167\phm{a} &   1.7\phm{A} &  0.0038  &                \\
            \phm{a} HD 39587 &        5136 (12) &    1.85 (0.02) &      0.45 (0.01) &          111 (1) &         1463 (17) &   0.89  &   143\phm{a} &   5.9\phm{A} &  0.0024  &                \\
            \phm{a} HD 65430 &       3138 (342) &     1.11 (0.2) &      0.32 (0.02) &           77 (1) &        3267 (302) &   0.78  &  67.8\phm{a} &   4.0\phm{A} &  0.00038 &                \\
           \phm{a} HD 122742 &         3617 (7) &    6.41 (0.01) &     0.48 (0.001) &        183 (0.1) &          2030 (3) &   0.96  &   545\phm{a} &   5.3\phm{A} &   0.0670 &       1,2,3,4,5 \\
           \phm{a} HD 131511 &  125.396 (0.001) &   19.10 (0.01) &     0.51 (0.001) &        219 (0.1) &   203.407 (0.004) &   0.78  &   455\phm{a} &  0.52\phm{A} &   0.0580 &             6,7 \\
\enddata

\tablerefs{(1) Duquennoy \& Mayor 1991; (2) Martin et~al. 1998; (3) Blazit et~al. 1987; (4)  
Kamper 1987; (5) Wagman 1949; (6) Beavers \& Salzer 1983; (7) Kamper  
\& Lyons 1981; (8) Halbwachs et~al. 2000; (9) Mazeh et~al. 1996;  
(10) Oetiker et~al. 2001; (11) Mayor et~al. 1997; (12) McCarthy 1986;  
(13) Barbieri et~al. 1996; (14) Henry \& McCarthy 1993; (15) Marcy \& Moore 1989; (16) McCarthy  
\& Henry 1987; (17) Lippincott \& Borgman 1978 }  
\thispagestyle{empty}  
\end{deluxetable} 
 
\eject

\begin{deluxetable}{llrr}
\tablenum{4}
\tablecaption{Stars With Linear Trends}
\tabletypesize{\small}
\label{rv4}
\tablewidth{0pt}
\tablehead{
\colhead{Primary} & \colhead{Alternate} & \colhead{Slope} & \colhead{N} \\
\colhead{Name} & \colhead{Name} & \colhead{(m s$^{-1}$ per day)} & \colhead{Obs} \\
}
\startdata
\tableline
            \phm{a} HD 3770 &            \phm{aa} HIP 3169 &            0.41 (0.02) \phn &               13 \\
            \phm{a} HD 3795 &            \phm{aa} HIP 3185 &          0.363 (0.003) \phn &               24 \\
           \phm{a} HD 11909 &            \phm{aa} HIP 9110 &             23.7 (0.8) \phn &                4 \\
           \phm{a} HD 14802 &           \phm{aa} HIP 11072 &           -1.73 (0.02) \phn &                7 \\
           \phm{a} HD 17037 &           \phm{aa} HIP 12764 &             -1.5 (0.1) \phn &               14 \\
           \phm{a} HD 21017 &           \phm{aa} HIP 15861 &            -11.3 (0.3) \phn &                5 \\
           \phm{a} HD 28388 &           \phm{aa} HIP 20802 &             -2.7 (0.3) \phn &                6 \\
           \phm{a} HD 29461 &           \phm{aa} HIP 21654 &            0.58 (0.02) \phn &                9 \\
\enddata
\tablenotetext{a}{Exhibits some additional positive curvature}
\tablenotetext{b}{Exhibits some additional negative curvature}
\end{deluxetable}

\end{document}